\documentclass[twocolumn,superscriptaddress,amsmath,amssymb,aps,pra]{revtex4}
\usepackage{amsmath,amsthm,amssymb}
\usepackage{latexsym}
\usepackage{amscd,graphicx}
\usepackage[titletoc]{appendix}
\usepackage{epsfig}
\usepackage{epstopdf}
\usepackage[normalem]{ulem}
\usepackage[dvipsnames]{xcolor}
\usepackage[colorlinks=true,linkcolor=red,citecolor=red]{hyperref}

\usepackage{pdfpages}
\usepackage{subfigure}
\usepackage{tabularx}

\newcommand{\abs}[1]{| #1 |}

\begin{document}
\title {Strong and noise-tolerant entanglement in dissipative optomechanics}

\author {Jiaojiao Chen}
\affiliation {Department of Physics and Optoelectronic Engineering,Anhui University, Anhui 230000, China}
	\date{\today }

\author{Wei Xiong}
\altaffiliation{xiongweiphys@wzu.edu.cn}
\affiliation{Department of Physics, Wenzhou University, Zhejiang, 325035, China}
\affiliation{International Quantum Academy, Shenzhen,518048,China}
	
\author{Dong Wang}
\affiliation{Department of Physics and Optoelectronic Engineering,Anhui University, Anhui 230000, China}

\author{Liu Ye}
\altaffiliation{yeliu@ahu.edu.cn}
\affiliation{Department of Physics and Optoelectronic Engineering,Anhui University, Anhui 230000, China}	
	\begin{abstract}
		Macroscopic entanglement, as {a} critical quantum {resource} in quantum information science, has been extensively studied in coherent optomechanics over the past decades. However, entanglement in dissipative optomechanics, where the cavity linewidth {depends on} the position of the mechanical resonator, remains {largely} unexplored. In this work, we {investigate} quantum entanglement in a {dissipative optomechanical system} {realized by a Michelson-Sagnac interferometer with} a movable membrane. {This configuration enables the switching between} coherent and dissipative optomechanical couplings at will. With {experimentally feasible} parameters, we demonstrate that the steady-state mechanical displacement exhibits a nonlinear (linear) dependence on the driving power {under} coherent (dissipative) coupling. {Furthermore}, we show that {the} quantum entanglement generated {via} dissipative coupling is significantly stronger and more {robust to noise} than that generated {via} coherent coupling. When both coherent and dissipative couplings are simultaneously {present}, {the} entanglement is {weakened} due to quantum interference. Our {results indicate that} dissipative optomechanical coupling {can} be a promising {route for engineering} strong and {noise-resilient} quantum entanglement.		
	\end{abstract}
	\maketitle

\section{Introduction}

Coherent optomechanics, where the cavity frequency {depends on} the position of the mechanical resonator, has garnered {significant} interest both theoretically and experimentally over the past two decades~\cite{aspelmeyer2014cavity}. In such systems, various proposals have been put forward, including ground-state cooling~\cite{teufel2011sideband,groblacher2009demonstration,schliesser2008resolved}, bistability~\cite{dorsel1983optical,gozzini1985light} and tristability~\cite{xiong2016cross}, optomechanically induced transparency~\cite{Optomechanically2010,safavi2011electromagnetically,PhysRevLett.111.133601}, quantum squeezing~\cite{liao2011parametric,nunnenkamp2010cooling,lu2015steady,aggarwal2020room,purdy2013strong,luo2022quantum}, phase {transitions}~\cite{wang2024quantum,jager2019dynamical,mann2018nonequilibrium,xu2017observation}, microwave-optical conversion~\cite{PhysRevLett.117.123902,han2021microwave,sahu2022quantum,andrews2014bidirectional,chen2023optomechanical,xu2016nonreciprocal}, spin-optomechanical {interfaces}~\cite{PhysRevB.103.174106,Chen:21,peng2023strong,PhysRevA.107.033516,PhysRevLett.123.053601}, higher-order exceptional {points}~\cite{xiong2021higher,xiong2022higher}, and macroscopic quantum entanglement~\cite{vitali2007optomechanical,PhysRevLett.108.153604,PhysRevLett.110.233602,PhysRevLett.110.253601,PhysRevA.109.043512,PhysRevB.108.024105,liu2024tunable,lai2022noise,jiao2020nonreciprocal,jiao2022nonreciprocal,liu2024nonreciprocal,shang2024resonance,ockeloen2018stabilized}. 

Recently, \textit{dissipative} optomechanics was theoretically proposed in a superconducting microwave circuit~\cite{elste2009quantum}, where the cavity linewidth is a function of the mechanical displacement~\cite{elste2009quantum}, giving rise to a dissipative coupling between the mechanical and cavity modes. Later, the setup was extended to {Michelson-Sagnac interferometers (MSIs)} with a movable membrane~\cite{xuereb2011dissipative,tarabrin2013anomalous}, where coherent or dissipative optomechanical coupling {can be switched on and off at will} by tuning the position of the membrane or the {reflectivity amplitude} of the beam splitter (BS). Experimentally, dissipative optomechanics {has been} demonstrated in {MSIs}~\cite{sawadsky2015observation,friedrich2011laser}, {a microdisk resonator coupled to a nanomechanical waveguide}~\cite{li2009reactive}, {a photonic crystal split-beam nanocavity}~\cite{wu2014dissipative}, {graphene drums coupled to a high-Q microsphere}~\cite{cole2015evanescent}, and {a levitated sphere trapped by dual-beam optical tweezers}~\cite{kuang2023nonlinear}. Within dissipative optomechanics, diverse intriguing quantum phenomena, such as unconventional bistability~\cite{kyriienko2014optomechanics}, quantum squeezing~\cite{qu2015generating,huang2020mechanical}, enhanced ground-state cooling~\cite{huang2018improving,liu2023mechanical}, quantum sensing~\cite{huang2017robust}, negative-damping instability~\cite{elste2009quantum}, self-sustained oscillations~\cite{huang2018dissipative}, and phonon lasing~\cite{kuang2023nonlinear,zhang2022dissipative}, have been investigated. However, macroscopic quantum entanglement has {not yet been revealed in} dissipative optomechanical coupling, although it has been intensively studied in coherent optomechanics.

In this work, we investigate quantum entanglement in an {experimentally realizable} dissipative optomechanical system~\cite{sawadsky2015observation,friedrich2011laser}. The setup consists of a fixed perfect mirror and an {effectively movable mirror implemented using} a Michelson-Sagnac interferometer (MSI) with an integrated membrane. 
We begin by analyzing the steady-state mechanical displacement under both coherent and dissipative coupling. Notably, with coherent coupling, the mechanical displacement exhibits a {nonlinear relationship} with the driving power, consistent with conventional optomechanics~\cite{aspelmeyer2014cavity}. In contrast, under dissipative coupling, the displacement shows a {linear dependence} on the driving power. Furthermore, we explore quantum entanglement in the presence of both coherent and dissipative coupling. Interestingly, we find that entanglement induced by dissipative coupling is not only significantly stronger than that generated by coherent coupling but also {considerably} more robust against bath temperature. For entanglement with coherent optomechanical coupling, its survival temperature is approximately~$\sim 8.5$ K, while for entanglement with dissipative optomechanical coupling, it can reach up to~$\sim 25$ K. When both coherent and dissipative couplings are simultaneously considered, entanglement {is reduced} due to quantum interference. Our work provides a promising path to generate strong and {temperature-resilient} quantum entanglement in quantum systems with \textit{dissipative coupling}.

The rest paper is organized as follows. In Sec.~\ref{sec2}, we introduce the proposed model and its Hamiltonian. Then, the steady-state solutions of the system are investigated in Sec.~\ref{sec3}. In Sec.~\ref{sec4}, the dynamics of the fluctuation and covariance matrix are given. In Secs.~\ref{sec5}, quantum entanglement is investigated. Finally, a conclusion is given in Sec.~\ref{sec6}.

\section{The model and Hamiltonian}\label{sec2}	
	\begin{figure}
	\includegraphics[scale=0.45]{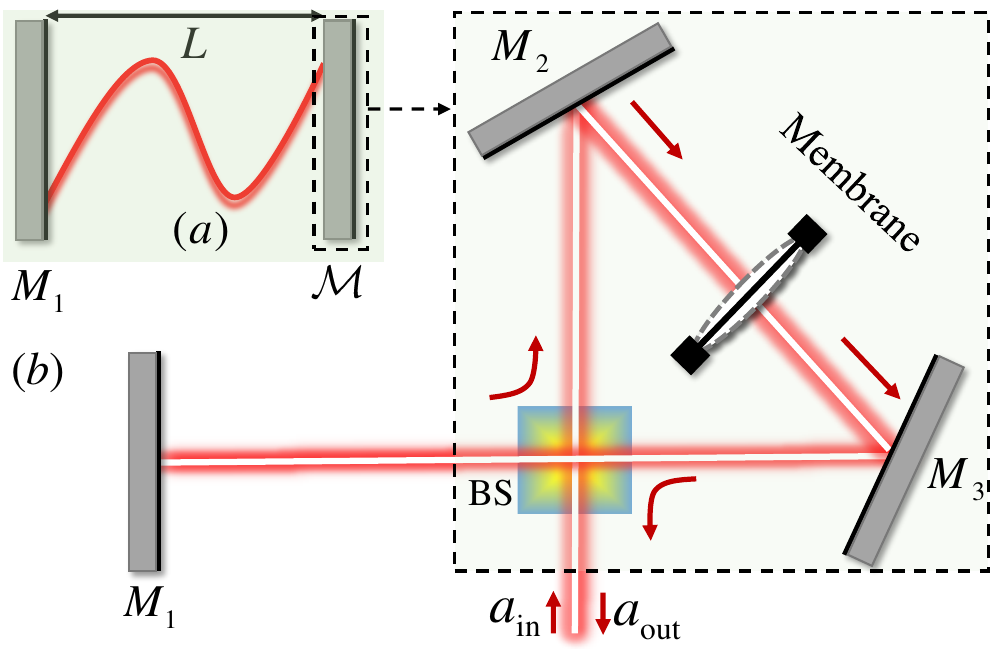}
	\caption{(a) Schematic diagram of a typical dissipative optomechanical system, consisting of a perfectly fixed mirror $M_1$ and an effective movable mirror $\mathcal{M}$.  
		(b) Structure of the effective movable mirror $\mathcal{M}$. It comprises a Michelson-Sagnac interferometer, which is formed by a beam splitter and two mirrors $M_2$ and $M_3$, along with a movable membrane (mechanical resonator). The parameter $L$ denotes the effective cavity length, where $2L$ is the round-trip length of the Sagnac mode ${\rm M}_1$-${\rm BS}$-${\rm M}_2$-${\rm M}_3$-${\rm BS}$. The optical field enters the system through the input port $a_{\rm in}$ and exits through the output port $a_{\rm out}$.	}
	\label{fig1}
\end{figure}	

As shown in Fig.~\ref{fig1}(a), we consider an experimentally feasible dissipative optomechanical system~\cite{sawadsky2015observation,friedrich2011laser}, consisting of a fixed mirror $M_1$ and an effective movable mirror $\mathcal{M}$. The effective mirror can be realized by a Michelson-Sagnac interferometer (MSI) {{incorporating} a movable membrane placed between mirrors $M_2$ and $M_3$ [see Fig.~\ref{fig1}(b)]. {{Here,} $L$ {denotes} the effective length of the cavity ({with $2L$ representing the round-trip length of the Sagnac mode} ${\rm M}_1$-${\rm BS}$-${\rm M}_2$-${\rm M}_3$-${\rm BS}$). The membrane is treated quantum mechanically, with its displacement and momentum operators denoted by $x$ and $p$, respectively, satisfying the canonical commutation relation $[x, p] = i\hbar$. Upon the introduction of an external driving field (with frequency $\omega_d$ and power $P$) at BS, the system can be effectively described as a typical dissipative optomechanical system, where both the cavity resonance frequency $\omega_a(x)$ and the decay rate $\kappa_a(x)$ depend on the membrane displacement $x$. {In the good-cavity limit~\cite{PhysRevA.81.033849}, }the system Hamiltonian is given by~\cite{elste2009quantum,xuereb2011dissipative,tarabrin2013anomalous} (setting $\hbar = 1$)
\begin{align}\label{eq1}
	H_S=&(\omega_a-g_\omega x)a^\dag a+\frac{1}{2}\omega_m (x^2+p^2)\notag\\
	&+\int d\omega\, \omega\, a_\omega^\dag a_\omega+i{\Gamma}\int \frac{d\omega}{\sqrt{2\pi}} (a_\omega 
	a^\dag-a_\omega^\dag a),
\end{align}
where $\omega_a$ ($\kappa_a$) denotes the cavity resonance frequency (decay rate) with annihilation (creation) operator $a$ ($a^\dag$) when the membrane is at equilibrium, $\omega_m$ is the mechanical resonance frequency of the membrane, and $\omega$ is the frequency of the bosonic bath mode with annihilation (creation) operator $a_\omega$ ($a_\omega^\dag$). The position-dependent coupling strength between the optical cavity and the bath is given by ${\Gamma}=\sqrt{2\kappa_a}-{g_\kappa x}/{\sqrt{2\kappa_a}}$. {Here, $g_{\omega(\kappa)}$ represents the single-photon coherent (dissipative) optomechanical coupling strength, explicitly expressed as~\cite{elste2009quantum,xuereb2011dissipative,tarabrin2013anomalous}
\begin{align}
	g_\omega=&-2(\omega_a x_{\rm zpf}/L)[(\abs{R}^2-\abs{T}^2)+\tau \cos(\arg \mathcal{T})],\notag\\ 
	g_\kappa=&-i\sqrt{2}(\omega_a x_{\rm zpf}/L)\abs{\tau}[2RT+\rho\cos(\arg \mathcal{T})],
\end{align}
with
\begin{align}
	\rho=&-[(R^2e^{2ix}+T^2e^{-2ix})\mathcal{R}+2RT\mathcal{T}]e^{-i \rm{arg}\mathcal{T}}, \notag\\
	\tau=&[(RT^*e^{2ix}-{\rm c.c.})\mathcal{R}-(\abs{R}^2-\abs{T}^2)\mathcal{T}]e^{-i \rm{arg}\mathcal{T}}
\end{align}
being the effective complex reflectivity and transmissivity of the mirror $\mathcal{M}$, which depend on the membrane's position $x$. Here, $x_{\rm zpf}$ denotes the zero-point fluctuation of the membrane, and $R~(\mathcal{R})$ and $T~(\mathcal{T})$ represent the complex reflectivity and transmissivity of BS (membrane), respectively. Therefore, the values of $g_\omega$ and $g_\kappa$ can be independently tuned, allowing one to switch them on or off as desired. In particular, a purely dissipative coupling (i.e., $g_\omega = 0$) can be achieved by ensuring $\abs{R} \neq \abs{T}$.}

Note that the bosonic mode $a_\omega$ in Eq.~(\ref{eq1}) is to introduce a zero-value expectation input field $a_{\rm in}=\int \frac{d\omega}{\sqrt{2\pi}}e^{-i\omega (t-t_0)} a_{\omega,0}$, with $a_{\omega,0}$ the operator at an initial time, to the cavity mode $a$, which can be easily obtained following the standard input-output theory with the Markov approximation. At the optical domain, the input field satisfies the relations~\cite{1998Quantum}
\begin{align}
	\langle a_{\rm in}(t)a_{\rm in}(t^\prime)\rangle=&0,\notag\\
	\langle a_{\rm in}(t)a^\dag_{\rm in}(t^\prime)\rangle=&\delta(t-t^\prime).
\end{align}
Substituting the input field $a_{\rm in}$ into Eq.~(\ref{eq1}), the Hamiltonian of the {\it driven} unconventional MSI, with respect to the driving frequency $\omega_d$, becomes
\begin{align}
	\mathcal{H}_S=&(\Delta_a-g_\omega x)a^\dag a+\frac{1}{2}\omega_m (x^2+p^2)\notag\\
	&+i{\Gamma} (A_{\rm in} a^\dag-A_{\rm in}^\dag a),
\end{align}
where $\Delta_a=\omega_a-\omega_d$ is the frequency detuning of the cavity mode from the driving field, and 
\begin{align}
	A_{\rm in}=a_{\rm in}+\mathcal{E}.
\end{align}
Here $\mathcal{E}=|\mathcal{E}|\exp(i\theta)$, with $|\mathcal{E}|=\sqrt{\mathcal{P}/\hbar\omega_d}$ being the amplitude of the driving field and $\theta$ the phase. Notably, $\mathcal{E}$ can also be interpreted as the steady-state value of the operator $A_{\rm in}$, while $a_{\rm in}$ represents the fluctuation of $A_{\rm in}$.

\section{Steady-state solution}\label{sec3}

When dissipation is included, the dynamics of the proposed system can be governed by the quantum Langevin equantion $\dot{\mathcal{O}}=-i[\mathcal{O},\mathcal{H}_{S}]+{ Dissipation}+{Noise}$, specifically,
\begin{align}\label{q3}
\dot{x}=&\omega_m p,\notag\\
\dot{p}=&g_\omega a^\dagger a-\omega_m x+\frac{ig_\kappa}{\sqrt{2\kappa_a}}(A_{\rm in}^\dag a-A_{\rm in} a^\dag)
-\gamma_m p+\xi,\notag\\
\dot{a}=&-(\kappa_a+i\Delta_a) a+ig_\omega ax+g_\kappa ax+{\Gamma} A_{\rm in},
\end{align}
where $\gamma_m$ is the damping rate of the membrane and $\xi$ is the Brownian noise operator arising from the coupling of the membrane oscillator to the hot environment. {Since the Brownian noise is inherently random and unbiased. In the statistical sense, its average value is zero, i.e., $\langle \xi\rangle=0$. However, investigating the fluctuation behavior of the system, the contribution from  the correlation function of $\xi$ should be taken into account, given by~\cite{giovannetti2001phase}}  $\langle\xi(t)\xi(t^\prime)\rangle=\frac{\gamma_m}{\omega_m}\int\frac{d\omega}{2\pi}e^{-i\omega(t-t^\prime)}\omega [{\rm coth}(\frac{\hbar\omega}{2k_b T})+1] $. In the limit of the high quality factor, $ \omega_m\gg\gamma_m $, the Brownian noise $\xi$ approximately becomes a  Markovian noise~\cite{benguria1981quantum}, i.e.,
\begin{align}\label{q4}
\langle\xi(t)\xi(t^\prime)+\xi(t^\prime)\xi(t)\rangle/2\backsimeq\gamma_m (2n_{\rm th}+1)\delta(t-t^\prime)
\end{align}
 where the mean thermal phonon number $ n_{\rm th}$ = $ [{\rm exp} (\hbar\omega_m/{k_B T}-1)]^{-1}$ with $k_B$ the Boltzmann constant and $T$ the bath temperature.

\begin{figure}
	\includegraphics[scale=0.5]{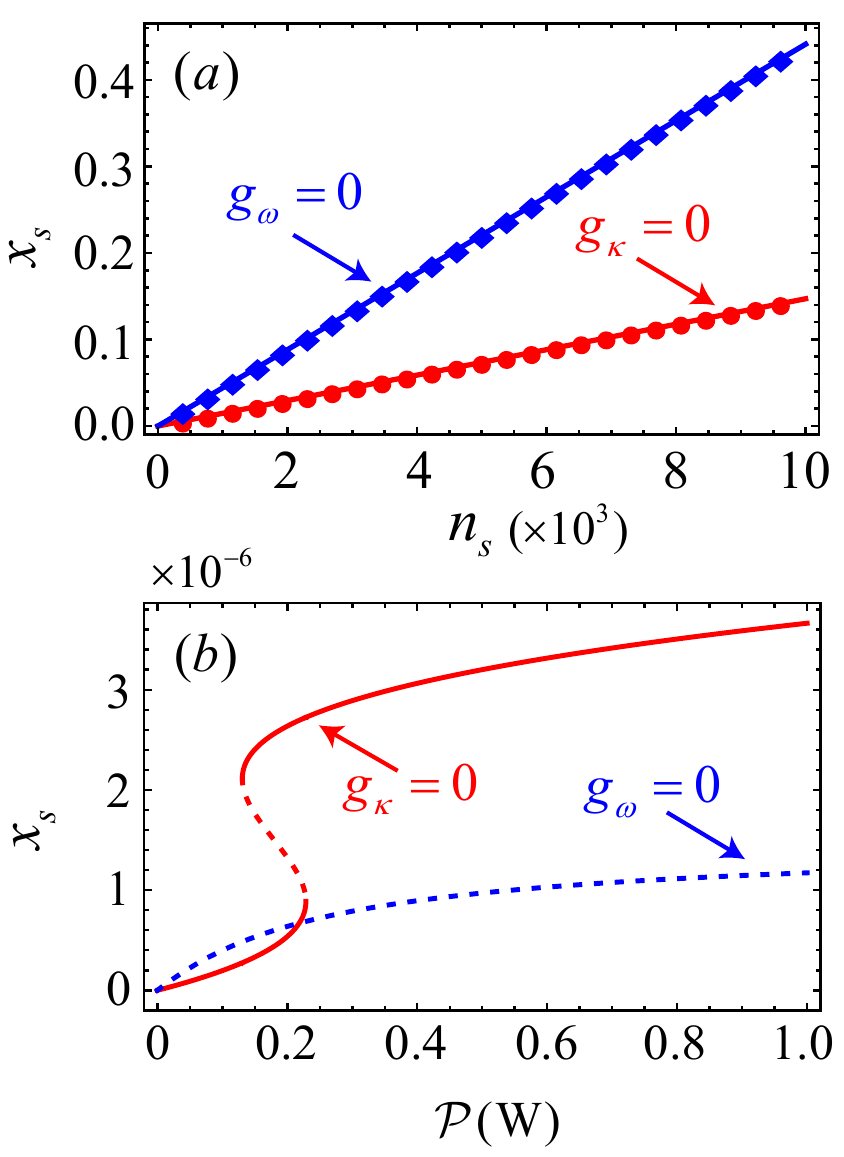}
	\caption{The normalized steady-state displacment $x_s$ of the membrane vs (a) the intracavity mean photon number $n_s=|a_s|^2$ and (b) the power of the driving field. The red curve corresponds to the case of the coherent coupling ($g_\omega/2\pi=2$ Hz). The blue curve is plotted for the dissipative coupling ($g_\kappa/2\pi=2$ Hz). In (a) and (b), $\Delta_a=3\kappa_a$ is assumed.}
	\label{fig2}
\end{figure}

{In the long-time limit ($t \rightarrow \infty$), the system reaches a steady state. Each operator $\mathcal{O}$ is then expanded as the sum of its classical steady-state value and a small quantum fluctuation, i.e., $\mathcal{O} = \mathcal{O}_s + \delta \mathcal{O}$. Substituting this decomposition into Eq.~(\ref{q3}), we derive a set of equations that govern the steady-state values $\mathcal{O}_s$ as follows:}
\begin{align}\label{eq5}
	\dot{x}_s=&\omega_m p_s,\notag\\
	\dot{p}_s=&g_\omega |a_s|^2-\omega_m x_s+\frac{ig_\kappa}{\sqrt{2\kappa_a}}(\mathcal{E}^* a_s-\mathcal{E} a_s^*)
	-\gamma_m p_s,\notag\\
	\dot{a}_s=&-(\kappa_s+i{\Delta}_s) a_s+{\Gamma_s}\mathcal{E},
\end{align}
where ${\Delta}_s=\Delta_a-{g_\omega} x_s$, $\kappa_s=\kappa_a-g_\kappa x_s$, and ${\Gamma_s}=\sqrt{2\kappa_a}-{g_\kappa x_s}/{\sqrt{2\kappa_a}}$.  At the steady state, we have $\dot{\mathcal{O}}_s=0$, {yielding to}
\begin{align}\label{eq6}
	p_s=&0,~~a_s=\frac{{\Gamma_s}\mathcal{E}}{\kappa_s+i{\Delta}_s},\notag\\
	x_s
	=&\frac{g_\omega n_s}{\omega_m}+\frac{i g_\kappa}{\sqrt{2\kappa_a}}\cdot\frac{a_s\mathcal{E}^*-a_s^*\mathcal{E}}{\omega_m}.
\end{align}
{Obviously,} the steady-state displacement of the membrane is determined by both the dispersive and dissipative couplings between photons and phonons. For \textit{coherent} coupling ($g_\kappa = 0$), $x_s$ is {\it proportional} to the intracavity mean photon number $n_s = |a_s|^2$ {[see the red curve in Fig.~\ref{fig2}(a)]}. For \textit{dissipative} coupling ($g_\omega = 0$), $x_s$ becomes
\begin{align}
	x_s=[1-\sqrt{1-2(g_\kappa/\kappa_a)^2(\Delta_a/\omega_m)n_s}]\omega_m\kappa_a,\label{eq10}
\end{align}
where $x_s\neq2\kappa_a/g_\kappa$ and $n_s\leq(\kappa_a/g_\kappa)^2(\omega_m/2\Delta_a)$ are both required. When $n_s=0$, $x_s=0$, i.e., the membrane is positioned at its equilibrum. {Here $x_s$ exhibits a quadratic dependence on $n_s$ in the dissipative coupling regime, as demonstrated by the blue curve in Fig.~\ref{fig2}(a).} {One can further observe} that the displacement of the membrane {induced} by coherent coupling is larger than that {induced} by dissipative coupling, as clearly shown in Fig.~\ref{fig2}(a). To plot Fig.~\ref{fig2}(a), we {employ} typical experimental parameters~\cite{sawadsky2015observation}: the wavelength of the driving field $\lambda = 1064$ nm, $\omega_d/2\pi = c/\lambda = 281.96$ THz, the quality factor of the membrane $\mathcal{Q}_m = 5.8 \times 10^5$, $\omega_m/2\pi = 136$ kHz, $\gamma_m/2\pi = 0.23$ Hz, $L = 8.7$ cm, $m = 80$ ng, and $\kappa_a/2\pi = 1.5$ MHz. {These parameters guarantee} the stability of the proposed system according to the Routh-Hurwitz criterion.

From Eq.~(\ref{eq6}), one can see that $x_s$ actually satisfies the following cubic equation related to the amplitude of the driving field $\mathcal{E}$,
\begin{align}\label{eq11}
	a x_s^3+bx_s^2+c x_s+d=0,
\end{align}
where
\begin{align}
	a=&2(g_\omega^2+g_\kappa^2)\omega_m\kappa_a,\notag\\
	b=&-4\omega_m\kappa_a(g_\omega\Delta_a +g_\kappa\kappa_a)-3g_\omega g_\kappa^2|\mathcal{E}|^2,\notag\\
	c=&2\omega_m\kappa_a(\kappa_a^2+\Delta_a^2)+2g_\kappa|\mathcal{E}|^2(g_\kappa\Delta_a+4g_\omega \kappa_a),\\
	d=&-4\kappa_a|\mathcal{E}|^2(g_\omega\kappa_a+g_\kappa\Delta_a)\notag.
\end{align}
{Specifically}, Eq.~(\ref{eq11}) may have three real roots, corresponding to bistability. {However}, bistability can only be predicted for coherent coupling, as shown by the red curve in Fig.~\ref{fig2}(b). {This suggests} that dissipative coupling {cannot} generate the typical bistability in optomechanics, as demonstrated by the blue curve in Fig.~\ref{fig2}(b).

\section{Covariance matrix}\label{sec4}

{Apart from the steady-state values, a set of equations related to the fluctuations $\delta \mathcal{O}$ can be obtained when the expression of $\mathcal{O} = \mathcal{O}_s + \delta \mathcal{O}$ is substituted back into Eq.~(\ref{q3}). Under} the strong driving field ($n_s \gg 1$), the higher-order fluctuation terms can be safely neglected~\cite{vitali2007optomechanical}, {which corresponds to the linearization of the system dynamics}. As a result, the dynamics for the fluctuations can be written as
\begin{align}\label{eq13}
	\dot{\delta x}=&\omega_m \delta p,\notag\\
	\dot{\delta p}=& \frac{G_\omega\delta a^\dagger+G_\omega^* \delta a}{\sqrt{2}}-\omega_m \delta x-\gamma_m\delta p+\xi\notag\\
	&+\frac{ig_\kappa(\mathcal{E}^*\delta a-\mathcal{E}\delta a^\dag)}{\sqrt{2\kappa_a}}+\frac{i(G_\kappa \delta a_{\rm in}^\dag-G_\kappa^*\delta a_{\rm in})}{2\sqrt{\kappa_a}},\\
	\dot{\delta a}=&-(\kappa_s+i\Delta_s) \delta a+(\frac{G_\kappa+iG_\omega}{\sqrt{2}} -\frac{g_\kappa\mathcal{E}}{\sqrt{2\kappa_a}})\delta x+{\Gamma_s} a_{\rm in},\notag
\end{align}
where $G_{\omega(\kappa)}=\sqrt{2}g_{\omega(\kappa)} a_s$ is the linearized dispersive (dissipative) coupling strength, enhanced by the amplitude of the driving field. By further defining
\begin{align}
	\delta x_a=&\frac{\delta a^\dag+\delta a}{\sqrt{2}},~~\delta p_a=i\frac{\delta a^\dag-\delta a}{\sqrt{2}},\notag\\
	x_{\rm in}=&\frac{a_{\rm in}^\dag+a_{\rm in}}{\sqrt{2}},~~p_{\rm in}=i\frac{a_{\rm in}^\dag-a_{\rm in}}{\sqrt{2}},
\end{align}
Eq.~(\ref{eq13}) can be rewritten in a more compact form, i.e.,
\begin{align}
	\dot{u}(t) = A u(t) + n(t),
\end{align}
where $u^T(t) = (\delta x, \delta p, \delta x_a, \delta p_a)$ is the vector of fluctuation operators, $ n(t) $ is the vector of noise operator, given by $n^T(t) = (0, \xi+\frac{G_\kappa}{\sqrt{2\kappa_a}}p_{\rm in}, {\Gamma_s} x_{\rm in}, {\Gamma_s} p_{\rm in})$, and the drift matrix $ A $ is
 \begin{equation}\label{q9}
 A=\left(\begin{array}{cccc}
 0  &\omega_m&0&0\\
 -\omega_m &-\gamma_m&G_\omega+\frac{G_\kappa\Delta_s }{\sqrt{2\kappa_a}{\Gamma_s}} &-\frac{G_\kappa\kappa_s}{\sqrt{2\kappa_a}{\Gamma_s}}\\
 G_\kappa-\frac{2G_\kappa\kappa_s}{\sqrt{2\kappa_a}{\Gamma_s}}  &0 &-\kappa_s&\Delta_s\\
 G_\omega-\frac{2G_\kappa\Delta_s }{\sqrt{2\kappa_a}{\Gamma_s}}&0&-\Delta_s&-\kappa_s\\
 \end{array}\right).
 \end{equation}
To obtain the drift matrix $A$, we have taken $a_s$ as real number, which can be realized by setting $\tan\theta=\Delta_a/\kappa_a$ in Eq.~(\ref{eq6}).

Since the phonon noise $\xi$ and the photon noise $a_{\rm in}$ are zero-mean quantum Gaussian noises and the dynamics is linearized [see Eq.~(\ref{eq13})], the quantum steady state for fluctuations is a zero-mean bipartite Gaussian state, fully described by a $4 \times 4$ correlation matrix ${V}$, where { $V_{ij}= \langle u_i(\infty) u_j(\infty) + u_j(\infty) u_i(\infty) \rangle / 2$ }with $i, j = 1, 2, 3, 4$. The $V$ is obtained by solving the Lyapunov equation $A V + V A^T = -D$, where $D$ is the diffusion matrix, defined by $\langle n_i(t) n_j(t') + n_j(t') n_i(t) \rangle / 2 = D_{ij} \delta(t - t')$. More specifically, $D$ can be expressed as
\begin{equation}\label{q10}
 D=\left(\begin{array}{cccc}
 0  &0&0&0\\
0 &\gamma_m(2n_{\rm th}+1)+\frac{G_\kappa^2}{4\kappa_a}&0 &\frac{G_\kappa}{\sqrt{2\kappa_a}}\frac{{\Gamma_s}}{2}\\
 0  &0 &\frac{{\Gamma_s^2}}{2}&0\\
 0&\frac{G_\kappa}{\sqrt{2\kappa_a}}\frac{{\Gamma_s}}{2}&0&\frac{{\Gamma_s^2}}{2}\\
 \end{array}\right).
\end{equation}
Compared to the coherent coupling case, dissipative coupling strength appears in the off-diagonal elements of the diffusion matrix, playing a {crucial} role in the generation of bipartite entanglement below.

To measure optomechanical entanglement, the logarithmic negativity ($E_N$) is taken~\cite{adesso2004extremal}, i.e.,
\begin{align}\label{q8}
E_N\equiv \rm max[0,-ln2\eta^-],
\end{align}
where $ \eta^-=2^{-1/2}[\Sigma({V})-(\Sigma({V})^2-4\rm det {{V}})^{1/2}]^{1/2}$ with $ \Sigma({V})=\rm{det} {A}+\rm{det} B-2\rm{det} C $,
${{V}}=\begin{pmatrix}
	A &C\\
	 {C^T}&  B\\
\end{pmatrix}
$ is the $ 2 \times2 $ block form. When $E_N>0~(=0)$, the state is entangled (separable). Moreover, the larger $E_N$, the stronger entanglement.

\section{Optomechanical Entanglement}\label{sec5}

To investigate optomechanical entanglement, we consider three specific scenarios: (i) coherent coupling, (ii) dissipative coupling, and (iii) {cooperative} coupling (i.e., both the coherent and dissipative couplings are considered).

\subsection{Coherent coupling ($g_\kappa = 0$)}

In the coherent coupling case ($g_\kappa = 0$), we present the entanglement measure $E_N$ as a function of both the coherent coupling strength $g_\omega$ and the normalized effective cavity detuning $\Delta_s/\omega_m$ in Fig.~\ref{fig3}(a) to explore the conditions for optimal optomechanical entanglement at a bath temperature of $T = 0.4$ K. The results reveal that the generated entanglement remains relatively weak compared to that typically observed in the conventional optomechanical system~\cite{vitali2007optomechanical}. {This is because a weaker coupling strength is necessary to maintain system stability, which in turn leads to weaker entanglement, while a stronger coupling strength can induce instability.} Nonetheless, optimal entanglement can still be achieved around the point ($\Delta_s/\omega_m, g_\omega/2\pi) \approx (2, 3.1)$, as marked by the green star, indicating that even under these constraints, entanglement can be optimized.

\begin{figure}
	\includegraphics[scale=0.4]{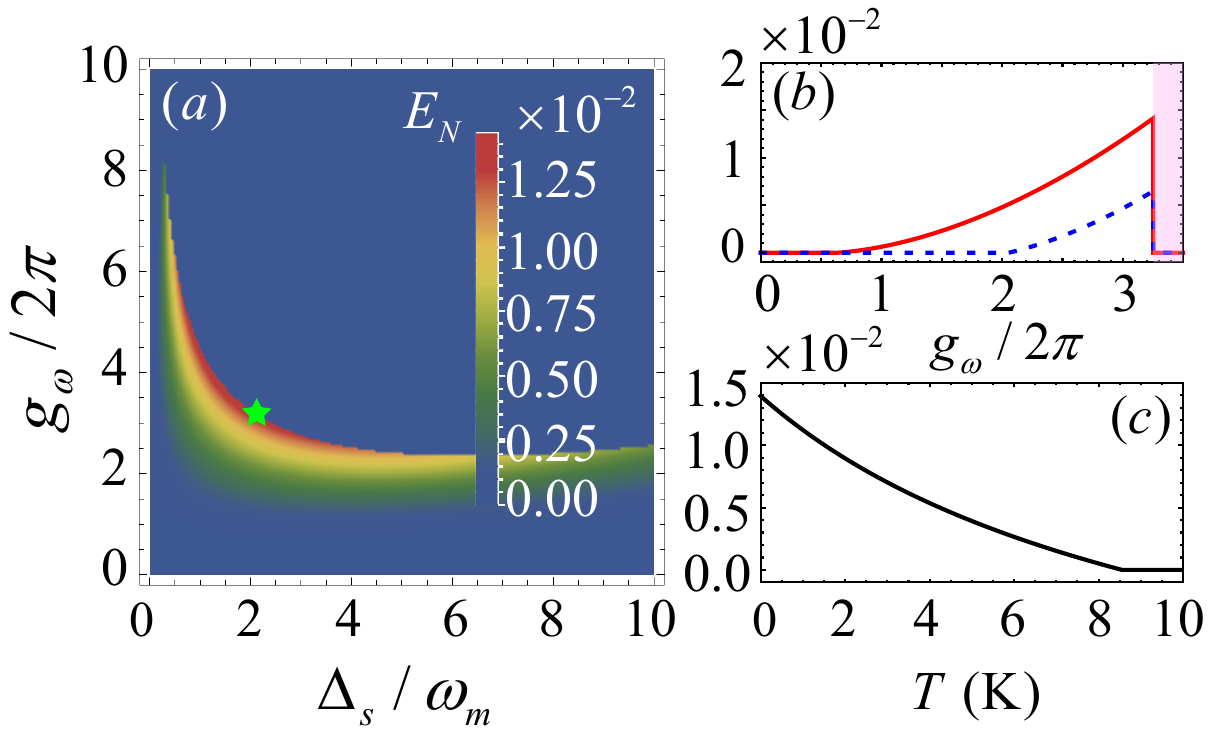}
	\caption{(a) The logarithmic negativity  $ E_N $ vs the normalized detuning $\Delta/\omega_m$ and the coherent coupling strength $g_\omega$ at bath temperature $T=400$ mK. (b) The logarithmic negativity  $ E_N $ vs the coherent coupling strength $g_\omega$ with $T=400$ mK (see the red curve) and $T=4$ K (see the blue dashed curve), where $\Delta_s/\omega_m=2$ is fixed. (c) The logarithmic negativity  $ E_N $ vs the bath temperature $T$ with $g_\omega/2\pi=3.1$ Hz and $\Delta_s/\omega_m=2$. Other parameters are the same as those in Fig.~\ref{fig2} except for $\mathcal{P}=50$ mW.}
	\label{fig3}
\end{figure}

To further investigate the coherent coupling effect on entanglement, we plot $E_N$ as a function of $g_\omega$ in Fig.~\ref{fig3}(b), where $\Delta_s \approx 2\omega_m$ is fixed. It is obviously shown that $E_N$ increases with $g_\omega$. Once up to its maximum,  $E_N$ will have a sharp reduction by further increasing $g_\omega$. Such abrupt behavior corresponds to the transition from stability to instability, as shown by the pink-shaded region. Additionally, when the bath temperature increases to $T = 4$ K [see the blue dashed curve], the coupling for generation of quantum entanglement is required to be larger and $E_N$ has a significant reduction. This indicates that quantum entanglement can be affected by the bath temperature. To clearly evaluate the effect of bath temperature, we plot $E_N$ as a function of the bath temperature in Fig.~\ref{fig3}(c), where $g_\omega/2\pi = 3.1$ Hz is fixed to obtain optimal entanglement. As the temperature increases, $E_N$ gradually declines. When the bath temperature increases to $T \approx 8.5$ K, $E_N=0$, {which indicates that quantum entanglement vanishes and the system becomes separable.}

\subsection{Dissipative coupling ($g_\omega = 0$)}

In dissipative coupling case ($g_\omega = 0$), we observe that the optimal entanglement can be significantly stronger than in the coherent coupling case, even surpassing the entanglement achieved in conventional optomechanical systems (i.e., $E_N = 0.35$)~\cite{vitali2007optomechanical}. This is demonstrated in Fig.~\ref{fig4}(a), where $E_N$ is plotted as a function of both the dissipative coupling strength ($g_\kappa$) and the normalized effective cavity detuning $\Delta_s/\omega_m$. The optimal entanglement is found at approximately ($\Delta_s/\omega_m, g_\kappa/2\pi) \approx (0.1, 20)$, as indicated by the green star. In contrast to the coherent coupling case [see Fig.~\ref{fig2}(a)], where the optimal entanglement occurs near the mechanical red sideband at $\Delta_s/\omega_m = 2$, the optimal entanglement in the dissipative coupling case is obtained when the cavity is nearly resonant with the driving field, specifically at $\Delta_s/\omega_m = 0.1$, which is far removed from the mechanical red sideband.

\begin{figure}
	\includegraphics[scale=0.4]{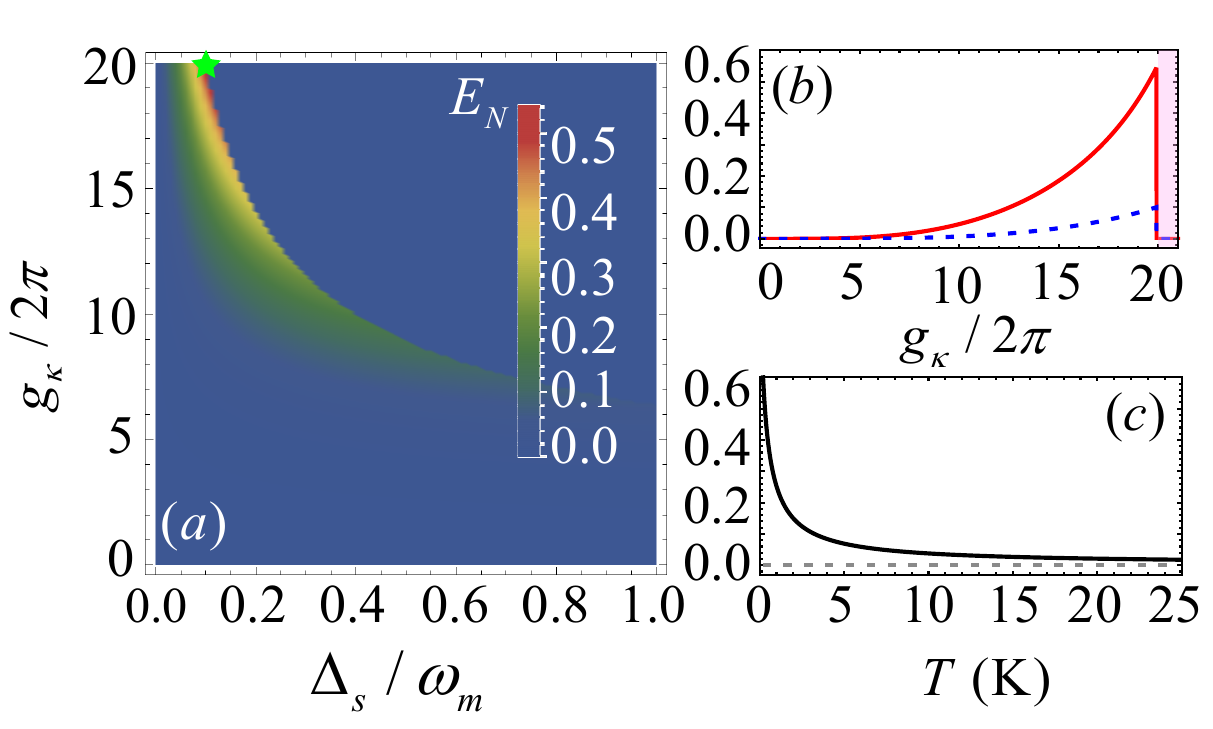}
	\caption{(a) The logarithmic negativity  $ E_N $ vs the normalized detuning $\Delta/\omega_m$ and the dissipative coupling strength $g_\kappa$ at bath temperature $T=400$ mK. (b) The logarithmic negativity  $ E_N $ vs the dissipative coupling strength $g_\kappa$ with $T=400$ mK (see the red curve) and $T=4$ K (see the blue dashed curve), where $\Delta_s/\omega_m=0.1$ is fixed. (c) The logarithmic negativity  $ E_N $ vs the bath temperature $T$ with $g_\kappa/2\pi=19$ Hz and $\Delta_s/\omega_m=0.1$. Other parameters are the same as those in Fig.~\ref{fig2} except for $\mathcal{P}=50$ mW.}
	\label{fig4}
\end{figure}

Furthermore, in Fig.~\ref{fig4}(b), we investigate the behavior of the optomechanical entanglement as a function of the dissipative coupling strength, while fixing $\Delta_s = 0.1\omega_m$. The resulting behavior of $E_N$ mirrors that observed in the coherent coupling case (see the red curve). The abrupt drop in entanglement marks the transition of the system from a stable to an unstable region, as indicated by the pink area. Notably, a relatively large dissipative coupling strength is required to achieve the optimal entanglement. This suggests that in the regime dominated by dissipative coupling, stronger interactions are necessary to generate significant quantum effects. As the bath temperature increases, for instance to $T = 4$ K, the entanglement is reduced (see the blue curve). To explore the influence of bath temperature on optomechanical entanglement, we fix {$g_\kappa/2\pi = 19$} Hz and present the results in Fig.~\ref{fig4}(c). We observe a sharp decrease in entanglement as the system temperature increases from~$0$ K to~$5$ K. However, beyond this point, the entanglement becomes more tolerant to further temperature increases, suggesting that the dissipative coupling becomes more robust to thermal noise at higher temperatures. {Remarkably, entanglement persists even up to~$25$ K. This indicates that strong, noise-resistant entanglement can be generated through dissipative optomechanical coupling, opening new ways for quantum information processing and communication in dissipative systems.}

\subsection{Cooperative coupling ($g_\omega \neq0$ and $g_\kappa \neq0$)}

\begin{figure}
	\includegraphics[scale=0.4]{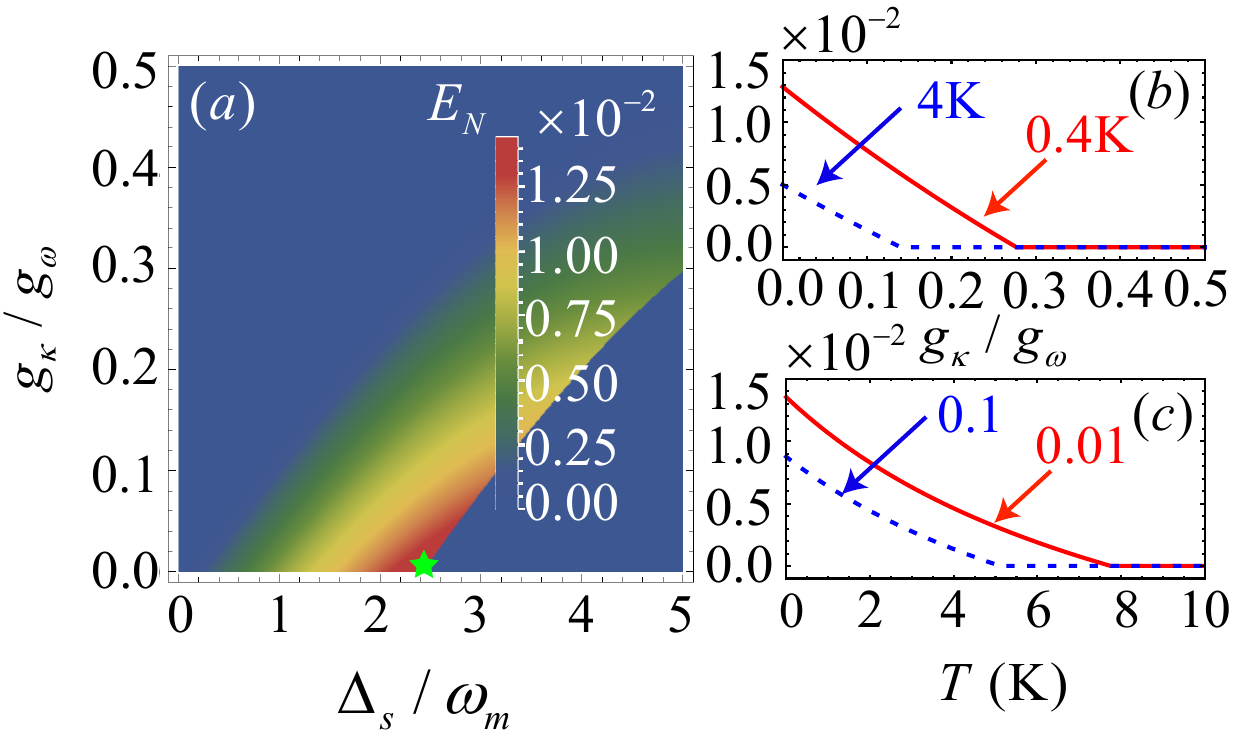}
	\caption{(a) The logarithmic negativity  $ E_N $ vs the normalized detuning $\Delta/\omega_m$ and the ratio of $g_\kappa/g_\omega$ at bath temperature $T=400$ mK. (b) The logarithmic negativity  $ E_N $ vs the ratio of $g_\kappa/g_\omega$ with $T=400$ mK (see the red curve) and $T=4$ K (see the blue dashed curve), where $\Delta_s/\omega_m=2.2$ is fixed. (c) The logarithmic negativity  $ E_N $ vs the bath temperature $T$ with { $g_\kappa/g_\omega=0.01$ (see the red curve) and $g_\kappa/g_\omega=0.1$ K (see the blue dashed curve)}, where $\Delta_s/\omega_m=2.2$. Other parameters are the same as those in Fig.~\ref{fig2} except for $\mathcal{P}=50$ mW, $g_\omega/2\pi=3$ Hz.}
	\label{fig5}
\end{figure}

To examine the effect of the cooperative coupling (i.e., coherent and dissipative couplings) on optomechanical entanglement, we set the coherent coupling strength to be $g_\omega/2\pi = 3$ Hz, corresponding to the optimal entanglement condition at $\Delta_s/\omega_m = 2$ in the dispersive coupling case [refer to Fig.~\ref{fig3}(a)]. By progressively increasing the dissipative coupling strength, we observe a gradual decline in the entanglement, as shown in Fig.~\ref{fig5}(a). {This degradation of entanglement results from the destructive interference between the coherent and dissipative coupling mechanisms.} When the bath temperature increases to $T = 4$ K, {the entanglement is further reduced}, as depicted by the blue dashed curve in Fig.~\ref{fig5}(b). {Furthermore, a larger} $g_\kappa$ {is found to be detrimental to the generation of entanglement.} This {is further illustrated} in Fig.~\ref{fig5}(c), where $g_\kappa/g_\omega = 0.01$ (the red curve) and $g_\kappa/g_\omega = 0.1$ (the blue curve) are {considered}. For $g_\kappa/g_\omega = 0.1$, the survival temperature of the entanglement is about~$5$ K, {whereas} for $g_\kappa/g_\omega = 0.01$, the survival temperature can reach up to~$8$ K.

\section{Conclusion}\label{sec6}

{Before concluding, we find it worthwhile to briefly discuss the conditions under which the coherent and dissipative coupling strengths, $g_\omega$ and $g_\kappa$, can become comparable or even equal, particularly in the context of implementing either purely coherent or purely dissipative coupling mechanisms.	In our setup, based on the Michelson-Sagnac interferometer configuration~\cite{xuereb2011dissipative,tarabrin2013anomalous,sawadsky2015observation,friedrich2011laser}, the single-photon coherent and dissipative coupling strengths can be independently tuned by adjusting the reflectivity of the beam splitter and the position of the membrane. This tunability allows access to both purely dispersive coupling ($g_\kappa = 0$) and purely dissipative coupling ($g_\omega = 0$). However, we emphasize that achieving equal values of $g_\omega$ and $g_\kappa$ under their respective coupling regimes involves different levels of experimental complexity. For instance, realizing a purely dissipative interaction typically demands more precise interferometric alignment and control compared to the conventional dispersive case. As a result, achieving equal values of $g_\omega$ in a purely coherent optomechanical system and $g_\kappa$ in a purely dissipative optomechanical system is generally challenging.	Fortunately, under strong driving conditions, the optomechanical system can be linearized, resulting in effective coupling strengths that are proportional to the amplitude of the driving field [see Eq.~(\ref{eq13})]. This enables the use of two independent MSI setups, one exhibiting purely dispersive coupling and the other purely dissipative coupling, each driven by an appropriately chosen strong optical field. Although the intrinsic coupling strengths $g_\omega$ and $g_\kappa$ may differ, by tailoring the driving field amplitudes, one can achieve comparable or even equal effective coupling strengths, $G_\omega$ and $G_\kappa$. This approach ensures a fair comparison in the investigation of quantum entanglement, as presented in Figs.~\ref{fig3} and \ref{fig4}, despite the inherent differences in the physical realization of the coupling mechanisms.}

In summary, we investigate quantum entanglement in a dissipative optomechanical system, realized by a fixed mirror and a Michelson-Sagnac interferometer with a membrane. This system allows the dissipative and coherent couplings to be independently tuned on and off by adjusting the membrane position or the reflectivity (transmissivity) of the beam splitter, enabling us to explore quantum entanglement in three distinct scenarios: coherent coupling, dissipative coupling, and cooperative coupling. {Our results reveal that, in the coherent coupling case, the steady-state mechanical displacement exhibits a nonlinear dependence on the driving power, while in the dissipative coupling case, the displacement shows a linear response.} Furthermore, entanglement generated through dissipative coupling is not only significantly stronger but also more noise-tolerant compared to that achieved via coherent coupling. {However, when both coupling mechanisms coexist, quantum entanglement is suppressed due to the destructive interference between the coherent and dissipative interactions.} These findings suggest that dissipative coupling {offers} a promising mechanism to engineer strong and noise-tolerant quantum entanglement.

This work is supported by the National Natural Science Foundation of China (Grant No.~12175001 ), the “Pioneer” and “Leading Goose” R$\&$D Program of Zhejiang (Grant No.~2025C01028), the Natural Science Foundation of Zhejiang Province (Grant No.~LY24A040004), and the Shenzhen International Quantum Academy (Grant No.~SIQA2024KFKT010).

\bibliography{ms2}

\begin{thebibliography}{68}
\expandafter\ifx\csname natexlab\endcsname\relax\def\natexlab#1{#1}\fi
\expandafter\ifx\csname bibnamefont\endcsname\relax
  \def\bibnamefont#1{#1}\fi
\expandafter\ifx\csname bibfnamefont\endcsname\relax
  \def\bibfnamefont#1{#1}\fi
\expandafter\ifx\csname citenamefont\endcsname\relax
  \def\citenamefont#1{#1}\fi
\expandafter\ifx\csname url\endcsname\relax
  \def\url#1{\texttt{#1}}\fi
\expandafter\ifx\csname urlprefix\endcsname\relax\def\urlprefix{URL }\fi
\providecommand{\bibinfo}[2]{#2}
\providecommand{\eprint}[2][]{\url{#2}}

\bibitem[{\citenamefont{Aspelmeyer et~al.}(2014)\citenamefont{Aspelmeyer,
  Kippenberg, and Marquardt}}]{aspelmeyer2014cavity}
\bibinfo{author}{\bibfnamefont{M.}~\bibnamefont{Aspelmeyer}},
  \bibinfo{author}{\bibfnamefont{T.~J.} \bibnamefont{Kippenberg}},
  \bibnamefont{and}
  \bibinfo{author}{\bibfnamefont{F.}~\bibnamefont{Marquardt}},
  \bibinfo{journal}{Rev. Mod. Phys.} \textbf{\bibinfo{volume}{86}},
  \bibinfo{pages}{1391} (\bibinfo{year}{2014}).

\bibitem[{\citenamefont{Teufel et~al.}(2011)\citenamefont{Teufel, Donner, Li,
  Harlow, Allman, Cicak, Sirois, Whittaker, Lehnert, and
  Simmonds}}]{teufel2011sideband}
\bibinfo{author}{\bibfnamefont{J.~D.} \bibnamefont{Teufel}},
  \bibinfo{author}{\bibfnamefont{T.}~\bibnamefont{Donner}},
  \bibinfo{author}{\bibfnamefont{D.}~\bibnamefont{Li}},
  \bibinfo{author}{\bibfnamefont{J.~W.} \bibnamefont{Harlow}},
  \bibinfo{author}{\bibfnamefont{M.}~\bibnamefont{Allman}},
  \bibinfo{author}{\bibfnamefont{K.}~\bibnamefont{Cicak}},
  \bibinfo{author}{\bibfnamefont{A.~J.} \bibnamefont{Sirois}},
  \bibinfo{author}{\bibfnamefont{J.~D.} \bibnamefont{Whittaker}},
  \bibinfo{author}{\bibfnamefont{K.~W.} \bibnamefont{Lehnert}},
  \bibnamefont{and} \bibinfo{author}{\bibfnamefont{R.~W.}
  \bibnamefont{Simmonds}}, \bibinfo{journal}{Nature}
  \textbf{\bibinfo{volume}{475}}, \bibinfo{pages}{359} (\bibinfo{year}{2011}).

\bibitem[{\citenamefont{Gr{\"o}blacher
  et~al.}(2009)\citenamefont{Gr{\"o}blacher, Hertzberg, Vanner, Cole, Gigan,
  Schwab, and Aspelmeyer}}]{groblacher2009demonstration}
\bibinfo{author}{\bibfnamefont{S.}~\bibnamefont{Gr{\"o}blacher}},
  \bibinfo{author}{\bibfnamefont{J.~B.} \bibnamefont{Hertzberg}},
  \bibinfo{author}{\bibfnamefont{M.~R.} \bibnamefont{Vanner}},
  \bibinfo{author}{\bibfnamefont{G.~D.} \bibnamefont{Cole}},
  \bibinfo{author}{\bibfnamefont{S.}~\bibnamefont{Gigan}},
  \bibinfo{author}{\bibfnamefont{K.}~\bibnamefont{Schwab}}, \bibnamefont{and}
  \bibinfo{author}{\bibfnamefont{M.}~\bibnamefont{Aspelmeyer}},
  \bibinfo{journal}{Nat. Phys.} \textbf{\bibinfo{volume}{5}},
  \bibinfo{pages}{485} (\bibinfo{year}{2009}).

\bibitem[{\citenamefont{Schliesser et~al.}(2008)\citenamefont{Schliesser,
  Rivi{\`e}re, Anetsberger, Arcizet, and Kippenberg}}]{schliesser2008resolved}
\bibinfo{author}{\bibfnamefont{A.}~\bibnamefont{Schliesser}},
  \bibinfo{author}{\bibfnamefont{R.}~\bibnamefont{Rivi{\`e}re}},
  \bibinfo{author}{\bibfnamefont{G.}~\bibnamefont{Anetsberger}},
  \bibinfo{author}{\bibfnamefont{O.}~\bibnamefont{Arcizet}}, \bibnamefont{and}
  \bibinfo{author}{\bibfnamefont{T.~J.} \bibnamefont{Kippenberg}},
  \bibinfo{journal}{Nat. Phys.} \textbf{\bibinfo{volume}{4}},
  \bibinfo{pages}{415} (\bibinfo{year}{2008}).

\bibitem[{\citenamefont{Dorsel et~al.}(1983)\citenamefont{Dorsel, McCullen,
  Meystre, Vignes, and Walther}}]{dorsel1983optical}
\bibinfo{author}{\bibfnamefont{A.}~\bibnamefont{Dorsel}},
  \bibinfo{author}{\bibfnamefont{J.~D.} \bibnamefont{McCullen}},
  \bibinfo{author}{\bibfnamefont{P.}~\bibnamefont{Meystre}},
  \bibinfo{author}{\bibfnamefont{E.}~\bibnamefont{Vignes}}, \bibnamefont{and}
  \bibinfo{author}{\bibfnamefont{H.}~\bibnamefont{Walther}},
  \bibinfo{journal}{Phys. Rev. Lett.} \textbf{\bibinfo{volume}{51}},
  \bibinfo{pages}{1550} (\bibinfo{year}{1983}).

\bibitem[{\citenamefont{Gozzini et~al.}(1985)\citenamefont{Gozzini, Maccarrone,
  Mango, Longo, and Barbarino}}]{gozzini1985light}
\bibinfo{author}{\bibfnamefont{A.}~\bibnamefont{Gozzini}},
  \bibinfo{author}{\bibfnamefont{F.}~\bibnamefont{Maccarrone}},
  \bibinfo{author}{\bibfnamefont{F.}~\bibnamefont{Mango}},
  \bibinfo{author}{\bibfnamefont{I.}~\bibnamefont{Longo}}, \bibnamefont{and}
  \bibinfo{author}{\bibfnamefont{S.}~\bibnamefont{Barbarino}},
  \bibinfo{journal}{JOSA B} \textbf{\bibinfo{volume}{2}}, \bibinfo{pages}{1841}
  (\bibinfo{year}{1985}).

\bibitem[{\citenamefont{Xiong et~al.}(2016)\citenamefont{Xiong, Jin, Qiu, Lam,
  and You}}]{xiong2016cross}
\bibinfo{author}{\bibfnamefont{W.}~\bibnamefont{Xiong}},
  \bibinfo{author}{\bibfnamefont{D.-Y.} \bibnamefont{Jin}},
  \bibinfo{author}{\bibfnamefont{Y.}~\bibnamefont{Qiu}},
  \bibinfo{author}{\bibfnamefont{C.-H.} \bibnamefont{Lam}}, \bibnamefont{and}
  \bibinfo{author}{\bibfnamefont{J.}~\bibnamefont{You}},
  \bibinfo{journal}{Phys. Rev. A} \textbf{\bibinfo{volume}{93}},
  \bibinfo{pages}{023844} (\bibinfo{year}{2016}).

\bibitem[{\citenamefont{Weis et~al.}(2010)\citenamefont{Weis, Rivière,
  Deléglise, Gavartin, Arcizet, Schliesser, and
  Kippenberg}}]{Optomechanically2010}
\bibinfo{author}{\bibfnamefont{S.}~\bibnamefont{Weis}},
  \bibinfo{author}{\bibfnamefont{R.}~\bibnamefont{Rivière}},
  \bibinfo{author}{\bibfnamefont{S.}~\bibnamefont{Deléglise}},
  \bibinfo{author}{\bibfnamefont{E.}~\bibnamefont{Gavartin}},
  \bibinfo{author}{\bibfnamefont{O.}~\bibnamefont{Arcizet}},
  \bibinfo{author}{\bibfnamefont{A.}~\bibnamefont{Schliesser}},
  \bibnamefont{and} \bibinfo{author}{\bibfnamefont{T.~J.}
  \bibnamefont{Kippenberg}}, \bibinfo{journal}{Science}
  \textbf{\bibinfo{volume}{330}}, \bibinfo{pages}{1520} (\bibinfo{year}{2010}).

\bibitem[{\citenamefont{Safavi-Naeini et~al.}(2011)\citenamefont{Safavi-Naeini,
  Alegre, Chan, Eichenfield, Winger, Lin, Hill, Chang, and
  Painter}}]{safavi2011electromagnetically}
\bibinfo{author}{\bibfnamefont{A.~H.} \bibnamefont{Safavi-Naeini}},
  \bibinfo{author}{\bibfnamefont{T.~M.} \bibnamefont{Alegre}},
  \bibinfo{author}{\bibfnamefont{J.}~\bibnamefont{Chan}},
  \bibinfo{author}{\bibfnamefont{M.}~\bibnamefont{Eichenfield}},
  \bibinfo{author}{\bibfnamefont{M.}~\bibnamefont{Winger}},
  \bibinfo{author}{\bibfnamefont{Q.}~\bibnamefont{Lin}},
  \bibinfo{author}{\bibfnamefont{J.~T.} \bibnamefont{Hill}},
  \bibinfo{author}{\bibfnamefont{D.~E.} \bibnamefont{Chang}}, \bibnamefont{and}
  \bibinfo{author}{\bibfnamefont{O.}~\bibnamefont{Painter}},
  \bibinfo{journal}{Nature} \textbf{\bibinfo{volume}{472}}, \bibinfo{pages}{69}
  (\bibinfo{year}{2011}).

\bibitem[{\citenamefont{Kronwald and Marquardt}(2013)}]{PhysRevLett.111.133601}
\bibinfo{author}{\bibfnamefont{A.}~\bibnamefont{Kronwald}} \bibnamefont{and}
  \bibinfo{author}{\bibfnamefont{F.}~\bibnamefont{Marquardt}},
  \bibinfo{journal}{Phys. Rev. Lett.} \textbf{\bibinfo{volume}{111}},
  \bibinfo{pages}{133601} (\bibinfo{year}{2013}).

\bibitem[{\citenamefont{Liao and Law}(2011)}]{liao2011parametric}
\bibinfo{author}{\bibfnamefont{J.-Q.} \bibnamefont{Liao}} \bibnamefont{and}
  \bibinfo{author}{\bibfnamefont{C.~K.} \bibnamefont{Law}},
  \bibinfo{journal}{Phys. Rev. A} \textbf{\bibinfo{volume}{83}},
  \bibinfo{pages}{033820} (\bibinfo{year}{2011}).

\bibitem[{\citenamefont{Nunnenkamp et~al.}(2010)\citenamefont{Nunnenkamp,
  B{\o}rkje, Harris, and Girvin}}]{nunnenkamp2010cooling}
\bibinfo{author}{\bibfnamefont{A.}~\bibnamefont{Nunnenkamp}},
  \bibinfo{author}{\bibfnamefont{K.}~\bibnamefont{B{\o}rkje}},
  \bibinfo{author}{\bibfnamefont{J.}~\bibnamefont{Harris}}, \bibnamefont{and}
  \bibinfo{author}{\bibfnamefont{S.}~\bibnamefont{Girvin}},
  \bibinfo{journal}{Phys. Rev. A} \textbf{\bibinfo{volume}{82}},
  \bibinfo{pages}{021806} (\bibinfo{year}{2010}).

\bibitem[{\citenamefont{L{\"u} et~al.}(2015)\citenamefont{L{\"u}, Liao, Tian,
  and Nori}}]{lu2015steady}
\bibinfo{author}{\bibfnamefont{X.-Y.} \bibnamefont{L{\"u}}},
  \bibinfo{author}{\bibfnamefont{J.-Q.} \bibnamefont{Liao}},
  \bibinfo{author}{\bibfnamefont{L.}~\bibnamefont{Tian}}, \bibnamefont{and}
  \bibinfo{author}{\bibfnamefont{F.}~\bibnamefont{Nori}},
  \bibinfo{journal}{Phys. Rev. A} \textbf{\bibinfo{volume}{91}},
  \bibinfo{pages}{013834} (\bibinfo{year}{2015}).

\bibitem[{\citenamefont{Aggarwal et~al.}(2020)\citenamefont{Aggarwal, Cullen,
  Cripe, Cole, Lanza, Libson, Follman, Heu, Corbitt, and
  Mavalvala}}]{aggarwal2020room}
\bibinfo{author}{\bibfnamefont{N.}~\bibnamefont{Aggarwal}},
  \bibinfo{author}{\bibfnamefont{T.~J.} \bibnamefont{Cullen}},
  \bibinfo{author}{\bibfnamefont{J.}~\bibnamefont{Cripe}},
  \bibinfo{author}{\bibfnamefont{G.~D.} \bibnamefont{Cole}},
  \bibinfo{author}{\bibfnamefont{R.}~\bibnamefont{Lanza}},
  \bibinfo{author}{\bibfnamefont{A.}~\bibnamefont{Libson}},
  \bibinfo{author}{\bibfnamefont{D.}~\bibnamefont{Follman}},
  \bibinfo{author}{\bibfnamefont{P.}~\bibnamefont{Heu}},
  \bibinfo{author}{\bibfnamefont{T.}~\bibnamefont{Corbitt}}, \bibnamefont{and}
  \bibinfo{author}{\bibfnamefont{N.}~\bibnamefont{Mavalvala}},
  \bibinfo{journal}{Nat. Phys.} \textbf{\bibinfo{volume}{16}},
  \bibinfo{pages}{784} (\bibinfo{year}{2020}).

\bibitem[{\citenamefont{Purdy et~al.}(2013)\citenamefont{Purdy, Yu, Peterson,
  Kampel, and Regal}}]{purdy2013strong}
\bibinfo{author}{\bibfnamefont{T.~P.} \bibnamefont{Purdy}},
  \bibinfo{author}{\bibfnamefont{P.-L.} \bibnamefont{Yu}},
  \bibinfo{author}{\bibfnamefont{R.~W.} \bibnamefont{Peterson}},
  \bibinfo{author}{\bibfnamefont{N.~S.} \bibnamefont{Kampel}},
  \bibnamefont{and} \bibinfo{author}{\bibfnamefont{C.~A.} \bibnamefont{Regal}},
  \bibinfo{journal}{Phys. Rev. X} \textbf{\bibinfo{volume}{3}},
  \bibinfo{pages}{031012} (\bibinfo{year}{2013}).

\bibitem[{\citenamefont{Luo et~al.}(2022)\citenamefont{Luo, Zhang, and
  Du}}]{luo2022quantum}
\bibinfo{author}{\bibfnamefont{X.-W.} \bibnamefont{Luo}},
  \bibinfo{author}{\bibfnamefont{C.}~\bibnamefont{Zhang}}, \bibnamefont{and}
  \bibinfo{author}{\bibfnamefont{S.}~\bibnamefont{Du}}, \bibinfo{journal}{Phys.
  Rev. Lett.} \textbf{\bibinfo{volume}{128}}, \bibinfo{pages}{173602}
  (\bibinfo{year}{2022}).

\bibitem[{\citenamefont{Wang et~al.}(2024)\citenamefont{Wang, Nori, and
  Xiang}}]{wang2024quantum}
\bibinfo{author}{\bibfnamefont{B.}~\bibnamefont{Wang}},
  \bibinfo{author}{\bibfnamefont{F.}~\bibnamefont{Nori}}, \bibnamefont{and}
  \bibinfo{author}{\bibfnamefont{Z.-L.} \bibnamefont{Xiang}},
  \bibinfo{journal}{Phys. Rev. Lett.} \textbf{\bibinfo{volume}{132}},
  \bibinfo{pages}{053601} (\bibinfo{year}{2024}).

\bibitem[{\citenamefont{J{\"a}ger et~al.}(2019)\citenamefont{J{\"a}ger, Cooper,
  Holland, and Morigi}}]{jager2019dynamical}
\bibinfo{author}{\bibfnamefont{S.~B.} \bibnamefont{J{\"a}ger}},
  \bibinfo{author}{\bibfnamefont{J.}~\bibnamefont{Cooper}},
  \bibinfo{author}{\bibfnamefont{M.~J.} \bibnamefont{Holland}},
  \bibnamefont{and} \bibinfo{author}{\bibfnamefont{G.}~\bibnamefont{Morigi}},
  \bibinfo{journal}{Phys. Rev. Lett.} \textbf{\bibinfo{volume}{123}},
  \bibinfo{pages}{053601} (\bibinfo{year}{2019}).

\bibitem[{\citenamefont{Mann et~al.}(2018)\citenamefont{Mann, Bakhtiari,
  Pelster, and Thorwart}}]{mann2018nonequilibrium}
\bibinfo{author}{\bibfnamefont{N.}~\bibnamefont{Mann}},
  \bibinfo{author}{\bibfnamefont{M.~R.} \bibnamefont{Bakhtiari}},
  \bibinfo{author}{\bibfnamefont{A.}~\bibnamefont{Pelster}}, \bibnamefont{and}
  \bibinfo{author}{\bibfnamefont{M.}~\bibnamefont{Thorwart}},
  \bibinfo{journal}{Phys. Rev. Lett.} \textbf{\bibinfo{volume}{120}},
  \bibinfo{pages}{063605} (\bibinfo{year}{2018}).

\bibitem[{\citenamefont{Xu et~al.}(2017)\citenamefont{Xu, Kemiktarak, Fan,
  Ragole, Lawall, and Taylor}}]{xu2017observation}
\bibinfo{author}{\bibfnamefont{H.}~\bibnamefont{Xu}},
  \bibinfo{author}{\bibfnamefont{U.}~\bibnamefont{Kemiktarak}},
  \bibinfo{author}{\bibfnamefont{J.}~\bibnamefont{Fan}},
  \bibinfo{author}{\bibfnamefont{S.}~\bibnamefont{Ragole}},
  \bibinfo{author}{\bibfnamefont{J.}~\bibnamefont{Lawall}}, \bibnamefont{and}
  \bibinfo{author}{\bibfnamefont{J.~M.} \bibnamefont{Taylor}},
  \bibinfo{journal}{Nat. Commun.} \textbf{\bibinfo{volume}{8}},
  \bibinfo{pages}{14481} (\bibinfo{year}{2017}).

\bibitem[{\citenamefont{Guo et~al.}(2016)\citenamefont{Guo, Zou, Jung, and
  Tang}}]{PhysRevLett.117.123902}
\bibinfo{author}{\bibfnamefont{X.}~\bibnamefont{Guo}},
  \bibinfo{author}{\bibfnamefont{C.-L.} \bibnamefont{Zou}},
  \bibinfo{author}{\bibfnamefont{H.}~\bibnamefont{Jung}}, \bibnamefont{and}
  \bibinfo{author}{\bibfnamefont{H.~X.} \bibnamefont{Tang}},
  \bibinfo{journal}{Phys. Rev. Lett.} \textbf{\bibinfo{volume}{117}},
  \bibinfo{pages}{123902} (\bibinfo{year}{2016}).

\bibitem[{\citenamefont{Han et~al.}(2021)\citenamefont{Han, Fu, Zou, Jiang, and
  Tang}}]{han2021microwave}
\bibinfo{author}{\bibfnamefont{X.}~\bibnamefont{Han}},
  \bibinfo{author}{\bibfnamefont{W.}~\bibnamefont{Fu}},
  \bibinfo{author}{\bibfnamefont{C.-L.} \bibnamefont{Zou}},
  \bibinfo{author}{\bibfnamefont{L.}~\bibnamefont{Jiang}}, \bibnamefont{and}
  \bibinfo{author}{\bibfnamefont{H.~X.} \bibnamefont{Tang}},
  \bibinfo{journal}{Optica} \textbf{\bibinfo{volume}{8}}, \bibinfo{pages}{1050}
  (\bibinfo{year}{2021}).

\bibitem[{\citenamefont{Sahu et~al.}(2022)\citenamefont{Sahu, Hease, Rueda,
  Arnold, Qiu, and Fink}}]{sahu2022quantum}
\bibinfo{author}{\bibfnamefont{R.}~\bibnamefont{Sahu}},
  \bibinfo{author}{\bibfnamefont{W.}~\bibnamefont{Hease}},
  \bibinfo{author}{\bibfnamefont{A.}~\bibnamefont{Rueda}},
  \bibinfo{author}{\bibfnamefont{G.}~\bibnamefont{Arnold}},
  \bibinfo{author}{\bibfnamefont{L.}~\bibnamefont{Qiu}}, \bibnamefont{and}
  \bibinfo{author}{\bibfnamefont{J.~M.} \bibnamefont{Fink}},
  \bibinfo{journal}{Nat. Commun.} \textbf{\bibinfo{volume}{13}},
  \bibinfo{pages}{1276} (\bibinfo{year}{2022}).

\bibitem[{\citenamefont{Andrews et~al.}(2014)\citenamefont{Andrews, Peterson,
  Purdy, Cicak, Simmonds, Regal, and Lehnert}}]{andrews2014bidirectional}
\bibinfo{author}{\bibfnamefont{R.~W.} \bibnamefont{Andrews}},
  \bibinfo{author}{\bibfnamefont{R.~W.} \bibnamefont{Peterson}},
  \bibinfo{author}{\bibfnamefont{T.~P.} \bibnamefont{Purdy}},
  \bibinfo{author}{\bibfnamefont{K.}~\bibnamefont{Cicak}},
  \bibinfo{author}{\bibfnamefont{R.~W.} \bibnamefont{Simmonds}},
  \bibinfo{author}{\bibfnamefont{C.~A.} \bibnamefont{Regal}}, \bibnamefont{and}
  \bibinfo{author}{\bibfnamefont{K.~W.} \bibnamefont{Lehnert}},
  \bibinfo{journal}{Nat. Phys.} \textbf{\bibinfo{volume}{10}},
  \bibinfo{pages}{321} (\bibinfo{year}{2014}).

\bibitem[{\citenamefont{Chen et~al.}(2023{\natexlab{a}})\citenamefont{Chen, Li,
  Lee, Chakravarthi, Fu, and Li}}]{chen2023optomechanical}
\bibinfo{author}{\bibfnamefont{I.-T.} \bibnamefont{Chen}},
  \bibinfo{author}{\bibfnamefont{B.}~\bibnamefont{Li}},
  \bibinfo{author}{\bibfnamefont{S.}~\bibnamefont{Lee}},
  \bibinfo{author}{\bibfnamefont{S.}~\bibnamefont{Chakravarthi}},
  \bibinfo{author}{\bibfnamefont{K.-M.} \bibnamefont{Fu}}, \bibnamefont{and}
  \bibinfo{author}{\bibfnamefont{M.}~\bibnamefont{Li}}, \bibinfo{journal}{Nat.
  Commun.} \textbf{\bibinfo{volume}{14}}, \bibinfo{pages}{7594}
  (\bibinfo{year}{2023}{\natexlab{a}}).

\bibitem[{\citenamefont{Xu et~al.}(2016)\citenamefont{Xu, Li, Chen, and
  Liu}}]{xu2016nonreciprocal}
\bibinfo{author}{\bibfnamefont{X.-W.} \bibnamefont{Xu}},
  \bibinfo{author}{\bibfnamefont{Y.}~\bibnamefont{Li}},
  \bibinfo{author}{\bibfnamefont{A.-X.} \bibnamefont{Chen}}, \bibnamefont{and}
  \bibinfo{author}{\bibfnamefont{Y.-x.} \bibnamefont{Liu}},
  \bibinfo{journal}{Phys. Rev. A} \textbf{\bibinfo{volume}{93}},
  \bibinfo{pages}{023827} (\bibinfo{year}{2016}).

\bibitem[{\citenamefont{Xiong et~al.}(2021{\natexlab{a}})\citenamefont{Xiong,
  Chen, Fang, Wang, Ye, and You}}]{PhysRevB.103.174106}
\bibinfo{author}{\bibfnamefont{W.}~\bibnamefont{Xiong}},
  \bibinfo{author}{\bibfnamefont{J.}~\bibnamefont{Chen}},
  \bibinfo{author}{\bibfnamefont{B.}~\bibnamefont{Fang}},
  \bibinfo{author}{\bibfnamefont{M.}~\bibnamefont{Wang}},
  \bibinfo{author}{\bibfnamefont{L.}~\bibnamefont{Ye}}, \bibnamefont{and}
  \bibinfo{author}{\bibfnamefont{J.~Q.} \bibnamefont{You}},
  \bibinfo{journal}{Phys. Rev. B} \textbf{\bibinfo{volume}{103}},
  \bibinfo{pages}{174106} (\bibinfo{year}{2021}{\natexlab{a}}).

\bibitem[{\citenamefont{Chen et~al.}(2021)\citenamefont{Chen, Li, Luo, Xiong,
  Wang, and Li}}]{Chen:21}
\bibinfo{author}{\bibfnamefont{J.}~\bibnamefont{Chen}},
  \bibinfo{author}{\bibfnamefont{Z.}~\bibnamefont{Li}},
  \bibinfo{author}{\bibfnamefont{X.-Q.} \bibnamefont{Luo}},
  \bibinfo{author}{\bibfnamefont{W.}~\bibnamefont{Xiong}},
  \bibinfo{author}{\bibfnamefont{M.}~\bibnamefont{Wang}}, \bibnamefont{and}
  \bibinfo{author}{\bibfnamefont{H.-C.} \bibnamefont{Li}},
  \bibinfo{journal}{Opt. Express} \textbf{\bibinfo{volume}{29}},
  \bibinfo{pages}{32639} (\bibinfo{year}{2021}).

\bibitem[{\citenamefont{Peng et~al.}(2023)\citenamefont{Peng, Tian, Chen,
  Zhang, Li, and Xiong}}]{peng2023strong}
\bibinfo{author}{\bibfnamefont{M.-L.} \bibnamefont{Peng}},
  \bibinfo{author}{\bibfnamefont{M.}~\bibnamefont{Tian}},
  \bibinfo{author}{\bibfnamefont{X.-C.} \bibnamefont{Chen}},
  \bibinfo{author}{\bibfnamefont{G.-Q.} \bibnamefont{Zhang}},
  \bibinfo{author}{\bibfnamefont{H.-C.} \bibnamefont{Li}}, \bibnamefont{and}
  \bibinfo{author}{\bibfnamefont{W.}~\bibnamefont{Xiong}},
  \bibinfo{journal}{arXiv preprint arXiv:2304.13553}  (\bibinfo{year}{2023}).

\bibitem[{\citenamefont{Xiong et~al.}(2023)\citenamefont{Xiong, Wang, Zhang,
  and Chen}}]{PhysRevA.107.033516}
\bibinfo{author}{\bibfnamefont{W.}~\bibnamefont{Xiong}},
  \bibinfo{author}{\bibfnamefont{M.}~\bibnamefont{Wang}},
  \bibinfo{author}{\bibfnamefont{G.-Q.} \bibnamefont{Zhang}}, \bibnamefont{and}
  \bibinfo{author}{\bibfnamefont{J.}~\bibnamefont{Chen}},
  \bibinfo{journal}{Phys. Rev. A} \textbf{\bibinfo{volume}{107}},
  \bibinfo{pages}{033516} (\bibinfo{year}{2023}).

\bibitem[{\citenamefont{J\"ager et~al.}(2019)\citenamefont{J\"ager, Cooper,
  Holland, and Morigi}}]{PhysRevLett.123.053601}
\bibinfo{author}{\bibfnamefont{S.~B.} \bibnamefont{J\"ager}},
  \bibinfo{author}{\bibfnamefont{J.}~\bibnamefont{Cooper}},
  \bibinfo{author}{\bibfnamefont{M.~J.} \bibnamefont{Holland}},
  \bibnamefont{and} \bibinfo{author}{\bibfnamefont{G.}~\bibnamefont{Morigi}},
  \bibinfo{journal}{Phys. Rev. Lett.} \textbf{\bibinfo{volume}{123}},
  \bibinfo{pages}{053601} (\bibinfo{year}{2019}).

\bibitem[{\citenamefont{Xiong et~al.}(2021{\natexlab{b}})\citenamefont{Xiong,
  Li, Song, Chen, Zhang, and Wang}}]{xiong2021higher}
\bibinfo{author}{\bibfnamefont{W.}~\bibnamefont{Xiong}},
  \bibinfo{author}{\bibfnamefont{Z.}~\bibnamefont{Li}},
  \bibinfo{author}{\bibfnamefont{Y.}~\bibnamefont{Song}},
  \bibinfo{author}{\bibfnamefont{J.}~\bibnamefont{Chen}},
  \bibinfo{author}{\bibfnamefont{G.-Q.} \bibnamefont{Zhang}}, \bibnamefont{and}
  \bibinfo{author}{\bibfnamefont{M.}~\bibnamefont{Wang}},
  \bibinfo{journal}{Phys. Rev. A} \textbf{\bibinfo{volume}{104}},
  \bibinfo{pages}{063508} (\bibinfo{year}{2021}{\natexlab{b}}).

\bibitem[{\citenamefont{Xiong et~al.}(2022)\citenamefont{Xiong, Li, Zhang,
  Wang, Li, Luo, and Chen}}]{xiong2022higher}
\bibinfo{author}{\bibfnamefont{W.}~\bibnamefont{Xiong}},
  \bibinfo{author}{\bibfnamefont{Z.}~\bibnamefont{Li}},
  \bibinfo{author}{\bibfnamefont{G.-Q.} \bibnamefont{Zhang}},
  \bibinfo{author}{\bibfnamefont{M.}~\bibnamefont{Wang}},
  \bibinfo{author}{\bibfnamefont{H.-C.} \bibnamefont{Li}},
  \bibinfo{author}{\bibfnamefont{X.-Q.} \bibnamefont{Luo}}, \bibnamefont{and}
  \bibinfo{author}{\bibfnamefont{J.}~\bibnamefont{Chen}},
  \bibinfo{journal}{Phys. Rev. A} \textbf{\bibinfo{volume}{106}},
  \bibinfo{pages}{033518} (\bibinfo{year}{2022}).

\bibitem[{\citenamefont{Vitali et~al.}(2007)\citenamefont{Vitali, Gigan,
  Ferreira, B{\"o}hm, Tombesi, Guerreiro, Vedral, Zeilinger, and
  Aspelmeyer}}]{vitali2007optomechanical}
\bibinfo{author}{\bibfnamefont{D.}~\bibnamefont{Vitali}},
  \bibinfo{author}{\bibfnamefont{S.}~\bibnamefont{Gigan}},
  \bibinfo{author}{\bibfnamefont{A.}~\bibnamefont{Ferreira}},
  \bibinfo{author}{\bibfnamefont{H.}~\bibnamefont{B{\"o}hm}},
  \bibinfo{author}{\bibfnamefont{P.}~\bibnamefont{Tombesi}},
  \bibinfo{author}{\bibfnamefont{A.}~\bibnamefont{Guerreiro}},
  \bibinfo{author}{\bibfnamefont{V.}~\bibnamefont{Vedral}},
  \bibinfo{author}{\bibfnamefont{A.}~\bibnamefont{Zeilinger}},
  \bibnamefont{and}
  \bibinfo{author}{\bibfnamefont{M.}~\bibnamefont{Aspelmeyer}},
  \bibinfo{journal}{Phys. Rev. Lett.} \textbf{\bibinfo{volume}{98}},
  \bibinfo{pages}{030405} (\bibinfo{year}{2007}).

\bibitem[{\citenamefont{Tian}(2012)}]{PhysRevLett.108.153604}
\bibinfo{author}{\bibfnamefont{L.}~\bibnamefont{Tian}}, \bibinfo{journal}{Phys.
  Rev. Lett.} \textbf{\bibinfo{volume}{108}}, \bibinfo{pages}{153604}
  (\bibinfo{year}{2012}).

\bibitem[{\citenamefont{Tian}(2013)}]{PhysRevLett.110.233602}
\bibinfo{author}{\bibfnamefont{L.}~\bibnamefont{Tian}}, \bibinfo{journal}{Phys.
  Rev. Lett.} \textbf{\bibinfo{volume}{110}}, \bibinfo{pages}{233602}
  (\bibinfo{year}{2013}).

\bibitem[{\citenamefont{Wang and Clerk}(2013)}]{PhysRevLett.110.253601}
\bibinfo{author}{\bibfnamefont{Y.-D.} \bibnamefont{Wang}} \bibnamefont{and}
  \bibinfo{author}{\bibfnamefont{A.~A.} \bibnamefont{Clerk}},
  \bibinfo{journal}{Phys. Rev. Lett.} \textbf{\bibinfo{volume}{110}},
  \bibinfo{pages}{253601} (\bibinfo{year}{2013}).

\bibitem[{\citenamefont{Chen et~al.}(2024)\citenamefont{Chen, Fan, Xiong, Wang,
  and Ye}}]{PhysRevA.109.043512}
\bibinfo{author}{\bibfnamefont{J.}~\bibnamefont{Chen}},
  \bibinfo{author}{\bibfnamefont{X.-G.} \bibnamefont{Fan}},
  \bibinfo{author}{\bibfnamefont{W.}~\bibnamefont{Xiong}},
  \bibinfo{author}{\bibfnamefont{D.}~\bibnamefont{Wang}}, \bibnamefont{and}
  \bibinfo{author}{\bibfnamefont{L.}~\bibnamefont{Ye}}, \bibinfo{journal}{Phys.
  Rev. A} \textbf{\bibinfo{volume}{109}}, \bibinfo{pages}{043512}
  (\bibinfo{year}{2024}).

\bibitem[{\citenamefont{Chen et~al.}(2023{\natexlab{b}})\citenamefont{Chen,
  Fan, Xiong, Wang, and Ye}}]{PhysRevB.108.024105}
\bibinfo{author}{\bibfnamefont{J.}~\bibnamefont{Chen}},
  \bibinfo{author}{\bibfnamefont{X.-G.} \bibnamefont{Fan}},
  \bibinfo{author}{\bibfnamefont{W.}~\bibnamefont{Xiong}},
  \bibinfo{author}{\bibfnamefont{D.}~\bibnamefont{Wang}}, \bibnamefont{and}
  \bibinfo{author}{\bibfnamefont{L.}~\bibnamefont{Ye}}, \bibinfo{journal}{Phys.
  Rev. B} \textbf{\bibinfo{volume}{108}}, \bibinfo{pages}{024105}
  (\bibinfo{year}{2023}{\natexlab{b}}).

\bibitem[{\citenamefont{Liu et~al.}(2024{\natexlab{a}})\citenamefont{Liu,
  Huang, and Xiong}}]{liu2024tunable}
\bibinfo{author}{\bibfnamefont{M.-Y.} \bibnamefont{Liu}},
  \bibinfo{author}{\bibfnamefont{X.-X.} \bibnamefont{Huang}}, \bibnamefont{and}
  \bibinfo{author}{\bibfnamefont{W.}~\bibnamefont{Xiong}}
  (\bibinfo{year}{2024}{\natexlab{a}}), \eprint{2404.15111}.

\bibitem[{\citenamefont{Lai et~al.}(2022)\citenamefont{Lai, Liao, Miranowicz,
  and Nori}}]{lai2022noise}
\bibinfo{author}{\bibfnamefont{D.-G.} \bibnamefont{Lai}},
  \bibinfo{author}{\bibfnamefont{J.-Q.} \bibnamefont{Liao}},
  \bibinfo{author}{\bibfnamefont{A.}~\bibnamefont{Miranowicz}},
  \bibnamefont{and} \bibinfo{author}{\bibfnamefont{F.}~\bibnamefont{Nori}},
  \bibinfo{journal}{Phys. Rev. Lett.} \textbf{\bibinfo{volume}{129}},
  \bibinfo{pages}{063602} (\bibinfo{year}{2022}).

\bibitem[{\citenamefont{Jiao et~al.}(2020)\citenamefont{Jiao, Zhang, Zhang,
  Miranowicz, Kuang, and Jing}}]{jiao2020nonreciprocal}
\bibinfo{author}{\bibfnamefont{Y.-F.} \bibnamefont{Jiao}},
  \bibinfo{author}{\bibfnamefont{S.-D.} \bibnamefont{Zhang}},
  \bibinfo{author}{\bibfnamefont{Y.-L.} \bibnamefont{Zhang}},
  \bibinfo{author}{\bibfnamefont{A.}~\bibnamefont{Miranowicz}},
  \bibinfo{author}{\bibfnamefont{L.-M.} \bibnamefont{Kuang}}, \bibnamefont{and}
  \bibinfo{author}{\bibfnamefont{H.}~\bibnamefont{Jing}},
  \bibinfo{journal}{Phys. Rev. Lett.} \textbf{\bibinfo{volume}{125}},
  \bibinfo{pages}{143605} (\bibinfo{year}{2020}).

\bibitem[{\citenamefont{Jiao et~al.}(2022)\citenamefont{Jiao, Liu, Li, Yang,
  Kuang, and Jing}}]{jiao2022nonreciprocal}
\bibinfo{author}{\bibfnamefont{Y.-F.} \bibnamefont{Jiao}},
  \bibinfo{author}{\bibfnamefont{J.-X.} \bibnamefont{Liu}},
  \bibinfo{author}{\bibfnamefont{Y.}~\bibnamefont{Li}},
  \bibinfo{author}{\bibfnamefont{R.}~\bibnamefont{Yang}},
  \bibinfo{author}{\bibfnamefont{L.-M.} \bibnamefont{Kuang}}, \bibnamefont{and}
  \bibinfo{author}{\bibfnamefont{H.}~\bibnamefont{Jing}},
  \bibinfo{journal}{Phys. Rev. Applied} \textbf{\bibinfo{volume}{18}},
  \bibinfo{pages}{064008} (\bibinfo{year}{2022}).

\bibitem[{\citenamefont{Liu et~al.}(2024{\natexlab{b}})\citenamefont{Liu, Gong,
  Chen, Wang, and Xiong}}]{liu2024nonreciprocal}
\bibinfo{author}{\bibfnamefont{M.-Y.} \bibnamefont{Liu}},
  \bibinfo{author}{\bibfnamefont{Y.}~\bibnamefont{Gong}},
  \bibinfo{author}{\bibfnamefont{J.}~\bibnamefont{Chen}},
  \bibinfo{author}{\bibfnamefont{Y.-W.} \bibnamefont{Wang}}, \bibnamefont{and}
  \bibinfo{author}{\bibfnamefont{W.}~\bibnamefont{Xiong}},
  \bibinfo{journal}{arXiv preprint arXiv:2412.20030}
  (\bibinfo{year}{2024}{\natexlab{b}}).

\bibitem[{\citenamefont{Shang and Li}(2024)}]{shang2024resonance}
\bibinfo{author}{\bibfnamefont{C.}~\bibnamefont{Shang}} \bibnamefont{and}
  \bibinfo{author}{\bibfnamefont{H.}~\bibnamefont{Li}}, \bibinfo{journal}{Phys.
  Rev. Appl.} \textbf{\bibinfo{volume}{21}}, \bibinfo{pages}{044048}
  (\bibinfo{year}{2024}).

\bibitem[{\citenamefont{Ockeloen-Korppi
  et~al.}(2018)\citenamefont{Ockeloen-Korppi, Damsk{\"a}gg, Pirkkalainen,
  Asjad, Clerk, Massel, Woolley, and
  Sillanp{\"a}{\"a}}}]{ockeloen2018stabilized}
\bibinfo{author}{\bibfnamefont{C.}~\bibnamefont{Ockeloen-Korppi}},
  \bibinfo{author}{\bibfnamefont{E.}~\bibnamefont{Damsk{\"a}gg}},
  \bibinfo{author}{\bibfnamefont{J.-M.} \bibnamefont{Pirkkalainen}},
  \bibinfo{author}{\bibfnamefont{M.}~\bibnamefont{Asjad}},
  \bibinfo{author}{\bibfnamefont{A.}~\bibnamefont{Clerk}},
  \bibinfo{author}{\bibfnamefont{F.}~\bibnamefont{Massel}},
  \bibinfo{author}{\bibfnamefont{M.}~\bibnamefont{Woolley}}, \bibnamefont{and}
  \bibinfo{author}{\bibfnamefont{M.}~\bibnamefont{Sillanp{\"a}{\"a}}},
  \bibinfo{journal}{Nature} \textbf{\bibinfo{volume}{556}},
  \bibinfo{pages}{478} (\bibinfo{year}{2018}).

\bibitem[{\citenamefont{Elste et~al.}(2009)\citenamefont{Elste, Girvin, and
  Clerk}}]{elste2009quantum}
\bibinfo{author}{\bibfnamefont{F.}~\bibnamefont{Elste}},
  \bibinfo{author}{\bibfnamefont{S.}~\bibnamefont{Girvin}}, \bibnamefont{and}
  \bibinfo{author}{\bibfnamefont{A.}~\bibnamefont{Clerk}},
  \bibinfo{journal}{Phys. Rev. Lett.} \textbf{\bibinfo{volume}{102}},
  \bibinfo{pages}{207209} (\bibinfo{year}{2009}).

\bibitem[{\citenamefont{Xuereb et~al.}(2011)\citenamefont{Xuereb, Schnabel, and
  Hammerer}}]{xuereb2011dissipative}
\bibinfo{author}{\bibfnamefont{A.}~\bibnamefont{Xuereb}},
  \bibinfo{author}{\bibfnamefont{R.}~\bibnamefont{Schnabel}}, \bibnamefont{and}
  \bibinfo{author}{\bibfnamefont{K.}~\bibnamefont{Hammerer}},
  \bibinfo{journal}{Phys. Rev. Lett.} \textbf{\bibinfo{volume}{107}},
  \bibinfo{pages}{213604} (\bibinfo{year}{2011}).

\bibitem[{\citenamefont{Tarabrin et~al.}(2013)\citenamefont{Tarabrin, Kaufer,
  Khalili, Schnabel, and Hammerer}}]{tarabrin2013anomalous}
\bibinfo{author}{\bibfnamefont{S.~P.} \bibnamefont{Tarabrin}},
  \bibinfo{author}{\bibfnamefont{H.}~\bibnamefont{Kaufer}},
  \bibinfo{author}{\bibfnamefont{F.~Y.} \bibnamefont{Khalili}},
  \bibinfo{author}{\bibfnamefont{R.}~\bibnamefont{Schnabel}}, \bibnamefont{and}
  \bibinfo{author}{\bibfnamefont{K.}~\bibnamefont{Hammerer}},
  \bibinfo{journal}{Phys. Rev. A} \textbf{\bibinfo{volume}{88}},
  \bibinfo{pages}{023809} (\bibinfo{year}{2013}).

\bibitem[{\citenamefont{Sawadsky et~al.}(2015)\citenamefont{Sawadsky, Kaufer,
  Nia, Tarabrin, Khalili, Hammerer, and Schnabel}}]{sawadsky2015observation}
\bibinfo{author}{\bibfnamefont{A.}~\bibnamefont{Sawadsky}},
  \bibinfo{author}{\bibfnamefont{H.}~\bibnamefont{Kaufer}},
  \bibinfo{author}{\bibfnamefont{R.~M.} \bibnamefont{Nia}},
  \bibinfo{author}{\bibfnamefont{S.~P.} \bibnamefont{Tarabrin}},
  \bibinfo{author}{\bibfnamefont{F.~Y.} \bibnamefont{Khalili}},
  \bibinfo{author}{\bibfnamefont{K.}~\bibnamefont{Hammerer}}, \bibnamefont{and}
  \bibinfo{author}{\bibfnamefont{R.}~\bibnamefont{Schnabel}},
  \bibinfo{journal}{Phys. Rev. Lett.} \textbf{\bibinfo{volume}{114}},
  \bibinfo{pages}{043601} (\bibinfo{year}{2015}).

\bibitem[{\citenamefont{Friedrich et~al.}(2011)\citenamefont{Friedrich, Kaufer,
  Westphal, Yamamoto, Sawadsky, Khalili, Danilishin, Go{\ss}ler, Danzmann, and
  Schnabel}}]{friedrich2011laser}
\bibinfo{author}{\bibfnamefont{D.}~\bibnamefont{Friedrich}},
  \bibinfo{author}{\bibfnamefont{H.}~\bibnamefont{Kaufer}},
  \bibinfo{author}{\bibfnamefont{T.}~\bibnamefont{Westphal}},
  \bibinfo{author}{\bibfnamefont{K.}~\bibnamefont{Yamamoto}},
  \bibinfo{author}{\bibfnamefont{A.}~\bibnamefont{Sawadsky}},
  \bibinfo{author}{\bibfnamefont{F.~Y.} \bibnamefont{Khalili}},
  \bibinfo{author}{\bibfnamefont{S.}~\bibnamefont{Danilishin}},
  \bibinfo{author}{\bibfnamefont{S.}~\bibnamefont{Go{\ss}ler}},
  \bibinfo{author}{\bibfnamefont{K.}~\bibnamefont{Danzmann}}, \bibnamefont{and}
  \bibinfo{author}{\bibfnamefont{R.}~\bibnamefont{Schnabel}},
  \bibinfo{journal}{New Journal of Physics} \textbf{\bibinfo{volume}{13}},
  \bibinfo{pages}{093017} (\bibinfo{year}{2011}).

\bibitem[{\citenamefont{Li et~al.}(2009)\citenamefont{Li, Pernice, and
  Tang}}]{li2009reactive}
\bibinfo{author}{\bibfnamefont{M.}~\bibnamefont{Li}},
  \bibinfo{author}{\bibfnamefont{W.~H.} \bibnamefont{Pernice}},
  \bibnamefont{and} \bibinfo{author}{\bibfnamefont{H.~X.} \bibnamefont{Tang}},
  \bibinfo{journal}{Phys. Rev. Lett.} \textbf{\bibinfo{volume}{103}},
  \bibinfo{pages}{223901} (\bibinfo{year}{2009}).

\bibitem[{\citenamefont{Wu et~al.}(2014)\citenamefont{Wu, Hryciw, Healey, Lake,
  Jayakumar, Freeman, Davis, and Barclay}}]{wu2014dissipative}
\bibinfo{author}{\bibfnamefont{M.}~\bibnamefont{Wu}},
  \bibinfo{author}{\bibfnamefont{A.~C.} \bibnamefont{Hryciw}},
  \bibinfo{author}{\bibfnamefont{C.}~\bibnamefont{Healey}},
  \bibinfo{author}{\bibfnamefont{D.~P.} \bibnamefont{Lake}},
  \bibinfo{author}{\bibfnamefont{H.}~\bibnamefont{Jayakumar}},
  \bibinfo{author}{\bibfnamefont{M.~R.} \bibnamefont{Freeman}},
  \bibinfo{author}{\bibfnamefont{J.~P.} \bibnamefont{Davis}}, \bibnamefont{and}
  \bibinfo{author}{\bibfnamefont{P.~E.} \bibnamefont{Barclay}},
  \bibinfo{journal}{Phys. Rev. X} \textbf{\bibinfo{volume}{4}},
  \bibinfo{pages}{021052} (\bibinfo{year}{2014}).

\bibitem[{\citenamefont{Cole et~al.}(2015)\citenamefont{Cole, Brawley, Adiga,
  De~Alba, Parpia, Ilic, Craighead, and Bowen}}]{cole2015evanescent}
\bibinfo{author}{\bibfnamefont{R.~M.} \bibnamefont{Cole}},
  \bibinfo{author}{\bibfnamefont{G.~A.} \bibnamefont{Brawley}},
  \bibinfo{author}{\bibfnamefont{V.~P.} \bibnamefont{Adiga}},
  \bibinfo{author}{\bibfnamefont{R.}~\bibnamefont{De~Alba}},
  \bibinfo{author}{\bibfnamefont{J.~M.} \bibnamefont{Parpia}},
  \bibinfo{author}{\bibfnamefont{B.}~\bibnamefont{Ilic}},
  \bibinfo{author}{\bibfnamefont{H.~G.} \bibnamefont{Craighead}},
  \bibnamefont{and} \bibinfo{author}{\bibfnamefont{W.~P.} \bibnamefont{Bowen}},
  \bibinfo{journal}{Phys. Rev. Applied} \textbf{\bibinfo{volume}{3}},
  \bibinfo{pages}{024004} (\bibinfo{year}{2015}).

\bibitem[{\citenamefont{Kuang et~al.}(2023)\citenamefont{Kuang, Huang, Xiong,
  Zuo, Han, Nori, Qiu, Luo, Jing, and Xiao}}]{kuang2023nonlinear}
\bibinfo{author}{\bibfnamefont{T.}~\bibnamefont{Kuang}},
  \bibinfo{author}{\bibfnamefont{R.}~\bibnamefont{Huang}},
  \bibinfo{author}{\bibfnamefont{W.}~\bibnamefont{Xiong}},
  \bibinfo{author}{\bibfnamefont{Y.}~\bibnamefont{Zuo}},
  \bibinfo{author}{\bibfnamefont{X.}~\bibnamefont{Han}},
  \bibinfo{author}{\bibfnamefont{F.}~\bibnamefont{Nori}},
  \bibinfo{author}{\bibfnamefont{C.-W.} \bibnamefont{Qiu}},
  \bibinfo{author}{\bibfnamefont{H.}~\bibnamefont{Luo}},
  \bibinfo{author}{\bibfnamefont{H.}~\bibnamefont{Jing}}, \bibnamefont{and}
  \bibinfo{author}{\bibfnamefont{G.}~\bibnamefont{Xiao}},
  \bibinfo{journal}{Nat. Phys.} \textbf{\bibinfo{volume}{19}},
  \bibinfo{pages}{414} (\bibinfo{year}{2023}).

\bibitem[{\citenamefont{Kyriienko et~al.}(2014)\citenamefont{Kyriienko, Liew,
  and Shelykh}}]{kyriienko2014optomechanics}
\bibinfo{author}{\bibfnamefont{O.}~\bibnamefont{Kyriienko}},
  \bibinfo{author}{\bibfnamefont{T.~C.~H.} \bibnamefont{Liew}},
  \bibnamefont{and} \bibinfo{author}{\bibfnamefont{I.~A.}
  \bibnamefont{Shelykh}}, \bibinfo{journal}{Phys. Rev. Lett.}
  \textbf{\bibinfo{volume}{112}}, \bibinfo{pages}{076402}
  (\bibinfo{year}{2014}).

\bibitem[{\citenamefont{Qu and Agarwal}(2015)}]{qu2015generating}
\bibinfo{author}{\bibfnamefont{K.}~\bibnamefont{Qu}} \bibnamefont{and}
  \bibinfo{author}{\bibfnamefont{G.}~\bibnamefont{Agarwal}},
  \bibinfo{journal}{Phys. Rev. A} \textbf{\bibinfo{volume}{91}},
  \bibinfo{pages}{063815} (\bibinfo{year}{2015}).

\bibitem[{\citenamefont{Huang and Chen}(2020)}]{huang2020mechanical}
\bibinfo{author}{\bibfnamefont{S.}~\bibnamefont{Huang}} \bibnamefont{and}
  \bibinfo{author}{\bibfnamefont{A.}~\bibnamefont{Chen}},
  \bibinfo{journal}{Phys. Rev. A} \textbf{\bibinfo{volume}{102}},
  \bibinfo{pages}{023503} (\bibinfo{year}{2020}).

\bibitem[{\citenamefont{Huang and Chen}(2018)}]{huang2018improving}
\bibinfo{author}{\bibfnamefont{S.}~\bibnamefont{Huang}} \bibnamefont{and}
  \bibinfo{author}{\bibfnamefont{A.}~\bibnamefont{Chen}},
  \bibinfo{journal}{Phys. Rev. A} \textbf{\bibinfo{volume}{98}},
  \bibinfo{pages}{063818} (\bibinfo{year}{2018}).

\bibitem[{\citenamefont{Liu et~al.}(2023)\citenamefont{Liu, Liu, Hu, Jiang, Wu,
  and Li}}]{liu2023mechanical}
\bibinfo{author}{\bibfnamefont{Y.}~\bibnamefont{Liu}},
  \bibinfo{author}{\bibfnamefont{Y.}~\bibnamefont{Liu}},
  \bibinfo{author}{\bibfnamefont{C.-S.} \bibnamefont{Hu}},
  \bibinfo{author}{\bibfnamefont{Y.-K.} \bibnamefont{Jiang}},
  \bibinfo{author}{\bibfnamefont{H.}~\bibnamefont{Wu}}, \bibnamefont{and}
  \bibinfo{author}{\bibfnamefont{Y.}~\bibnamefont{Li}}, \bibinfo{journal}{Phys.
  Rev. A} \textbf{\bibinfo{volume}{108}}, \bibinfo{pages}{023503}
  (\bibinfo{year}{2023}).

\bibitem[{\citenamefont{Huang and Agarwal}(2017)}]{huang2017robust}
\bibinfo{author}{\bibfnamefont{S.}~\bibnamefont{Huang}} \bibnamefont{and}
  \bibinfo{author}{\bibfnamefont{G.}~\bibnamefont{Agarwal}},
  \bibinfo{journal}{Phys. Rev. A} \textbf{\bibinfo{volume}{95}},
  \bibinfo{pages}{023844} (\bibinfo{year}{2017}).

\bibitem[{\citenamefont{Huang et~al.}(2018)\citenamefont{Huang, Li, Chin, Cai,
  Gu, Karim, Wu, Chen, Yang, Hao et~al.}}]{huang2018dissipative}
\bibinfo{author}{\bibfnamefont{J.}~\bibnamefont{Huang}},
  \bibinfo{author}{\bibfnamefont{Y.}~\bibnamefont{Li}},
  \bibinfo{author}{\bibfnamefont{L.~K.} \bibnamefont{Chin}},
  \bibinfo{author}{\bibfnamefont{H.}~\bibnamefont{Cai}},
  \bibinfo{author}{\bibfnamefont{Y.}~\bibnamefont{Gu}},
  \bibinfo{author}{\bibfnamefont{M.~F.} \bibnamefont{Karim}},
  \bibinfo{author}{\bibfnamefont{J.}~\bibnamefont{Wu}},
  \bibinfo{author}{\bibfnamefont{T.}~\bibnamefont{Chen}},
  \bibinfo{author}{\bibfnamefont{Z.}~\bibnamefont{Yang}},
  \bibinfo{author}{\bibfnamefont{Y.}~\bibnamefont{Hao}}, \bibnamefont{et~al.},
  \bibinfo{journal}{Appl. Phys. Lett.} \textbf{\bibinfo{volume}{112}}
  (\bibinfo{year}{2018}).

\bibitem[{\citenamefont{Zhang et~al.}(2022)\citenamefont{Zhang, Yang, Sheng,
  and Wu}}]{zhang2022dissipative}
\bibinfo{author}{\bibfnamefont{Q.}~\bibnamefont{Zhang}},
  \bibinfo{author}{\bibfnamefont{C.}~\bibnamefont{Yang}},
  \bibinfo{author}{\bibfnamefont{J.}~\bibnamefont{Sheng}}, \bibnamefont{and}
  \bibinfo{author}{\bibfnamefont{H.}~\bibnamefont{Wu}},
  \bibinfo{journal}{Proceedings of the National Academy of Sciences}
  \textbf{\bibinfo{volume}{119}}, \bibinfo{pages}{e2207543119}
  (\bibinfo{year}{2022}).

\bibitem[{\citenamefont{Yamamoto et~al.}(2010)\citenamefont{Yamamoto,
  Friedrich, Westphal, Go\ss{}ler, Danzmann, Somiya, Danilishin, and
  Schnabel}}]{PhysRevA.81.033849}
\bibinfo{author}{\bibfnamefont{K.}~\bibnamefont{Yamamoto}},
  \bibinfo{author}{\bibfnamefont{D.}~\bibnamefont{Friedrich}},
  \bibinfo{author}{\bibfnamefont{T.}~\bibnamefont{Westphal}},
  \bibinfo{author}{\bibfnamefont{S.}~\bibnamefont{Go\ss{}ler}},
  \bibinfo{author}{\bibfnamefont{K.}~\bibnamefont{Danzmann}},
  \bibinfo{author}{\bibfnamefont{K.}~\bibnamefont{Somiya}},
  \bibinfo{author}{\bibfnamefont{S.~L.} \bibnamefont{Danilishin}},
  \bibnamefont{and} \bibinfo{author}{\bibfnamefont{R.}~\bibnamefont{Schnabel}},
  \bibinfo{journal}{Phys. Rev. A} \textbf{\bibinfo{volume}{81}},
  \bibinfo{pages}{033849} (\bibinfo{year}{2010}).

\bibitem[{\citenamefont{Walls and Milburn}(1994)}]{1998Quantum}
\bibinfo{author}{\bibfnamefont{D.~F.} \bibnamefont{Walls}} \bibnamefont{and}
  \bibinfo{author}{\bibfnamefont{G.}~\bibnamefont{Milburn}},
  \emph{\bibinfo{title}{Quantum optics}} (\bibinfo{publisher}{Springer,
  Berlin}, \bibinfo{year}{1994}).

\bibitem[{\citenamefont{Giovannetti and Vitali}(2001)}]{giovannetti2001phase}
\bibinfo{author}{\bibfnamefont{V.}~\bibnamefont{Giovannetti}} \bibnamefont{and}
  \bibinfo{author}{\bibfnamefont{D.}~\bibnamefont{Vitali}},
  \bibinfo{journal}{Phys. Rev. A} \textbf{\bibinfo{volume}{63}},
  \bibinfo{pages}{023812} (\bibinfo{year}{2001}).

\bibitem[{\citenamefont{Benguria and Kac}(1981)}]{benguria1981quantum}
\bibinfo{author}{\bibfnamefont{R.}~\bibnamefont{Benguria}} \bibnamefont{and}
  \bibinfo{author}{\bibfnamefont{M.}~\bibnamefont{Kac}},
  \bibinfo{journal}{Phys. Rev. Lett.} \textbf{\bibinfo{volume}{46}},
  \bibinfo{pages}{1} (\bibinfo{year}{1981}).

\bibitem[{\citenamefont{Adesso et~al.}(2004)\citenamefont{Adesso, Serafini, and
  Illuminati}}]{adesso2004extremal}
\bibinfo{author}{\bibfnamefont{G.}~\bibnamefont{Adesso}},
  \bibinfo{author}{\bibfnamefont{A.}~\bibnamefont{Serafini}}, \bibnamefont{and}
  \bibinfo{author}{\bibfnamefont{F.}~\bibnamefont{Illuminati}},
  \bibinfo{journal}{Phys. Rev. A} \textbf{\bibinfo{volume}{70}},
  \bibinfo{pages}{022318} (\bibinfo{year}{2004}).

\end{thebibliography}
\end{document}